\documentclass{aa}
\usepackage{graphicx}
\usepackage{txfonts}
\usepackage{natbib}

\begin{document}

   \title{FIRST-based survey of Compact Steep Spectrum sources}

   \subtitle{II. MERLIN and VLA observations of Medium-sized Symmetric 
   Objects}

   \author{M. Kunert-Bajraszewska\inst{1}
          \and A. Marecki\inst{1}
          \and P. Thomasson\inst{2}
          \and R. E. Spencer\inst{2}
          }
   \authorrunning{Kunert-Bajraszewska et al.}
   \offprints{Andrzej Marecki\\ \email{amr@astro.uni.torun.pl}}
   \institute{Toru\'n Centre for Astronomy, N. Copernicus University,
              PL-87-100 Toru\'n, Poland
         \and
              Jodrell Bank Observatory, The University of Manchester, 
              Macclesfield, Cheshire, SK11 9DL, UK}

   \date{Received 7 December 2004 / Accepted 13 May 2005}

\abstract{A new sample of candidate Compact Steep Spectrum (CSS) sources that
are much weaker than the CSS source prototypes has been selected from the
VLA FIRST catalogue. MERLIN `snapshot' observations of the sources at 5\,GHz
indicate that six of them have an FR\,II-like morphology, but are not
edge-brightened as is normal for Medium-sized Symmetric Objects (MSOs) and
FR\,IIs. Further observations of these six sources with the VLA at 4.9\,GHz and
MERLIN at 1.7\,GHz, as well as subsequent full-track observations with MERLIN
at 5\,GHz of what appeared to be the two sources
of greatest interest are presented. The results are
discussed with reference to the established evolutionary model of CSS sources
being young but in which not all of them evolve to become old objects with
extended radio structures. A lack of stable fuelling in some of them may result
in an early transition to a so-called coasting phase so that they fade away
instead of growing to become large-scale objects. It is possible that one of the 
six sources (1542+323) could be labelled as a prematurely `dying' MSO or a `fader'.

}

\maketitle


\section{Introduction}

It is intriguing that the range of linear sizes of
FR\,II-type \citep{fr74} double-structured radio sources is very large and
spans from tens of parsecs to megaparsecs. For this and practical
reasons FR\,II-like radio sources have been roughly divided into three
classes: Compact Symmetric Objects (CSOs) --- those with Largest Angular Sizes
(LAS) below $1 h^{-1}$\,kpc, Medium-sized Symmetric Objects (MSOs) with
subgalactic sizes (LAS~$< 20 h^{-1}$\,kpc) and Large Symmetric Objects (LSOs)
--- LAS~$\ge 20h^{-1}$\,kpc \citep{fanti95}\footnote{For consistency with
earlier papers in this field, the following cosmological parameters have been
adopted throughout this paper:
$H_0$=100${\rm\,km\,s^{-1}\,Mpc^{-1}}$ and $q_0$=0.5.
Wherever in the text we refer to linear sizes we introduce $h^{-1}$.}.
Although the MSO class has been defined solely on morphological features
without particular stress on the spectral properties of its members
\citep{aug98}, many of the sources (if not the majority) have steep
spectra and consequently they have also been classified as Compact Steep
Spectrum (CSS) sources. It would also appear that MSOs (which are
unbeamed CSS sources) have similar morphologies to LSOs and that the lobe
expansion velocities in CSOs are high enough for the rapid evolution of CSOs
to MSOs \citep{oc98, ocp98}.

\citet{r96} proposed an evolutionary sequence unifying these three classes
of radio-loud AGNs (RLAGNs) having morphological similarity but with
different linear sizes. \citet{sn99, sn00} discussed many aspects of the
above scenario in detail. In particular, they concluded that the radio
luminosity of a CSO increases as it evolves, reaches a maximum in the MSO
phase and then gradually decreases as the object increases in size to
become
an LSO. The amount of time the source spends in each of these stages also
increases: CSOs are younger than $\sim$$10^4$~years, MSOs are typically
$\sim$$10^5$~years old and LSOs can manifest their activity for up to
$\sim$$10^8$~years. The ages of CSOs have been estimated from kinematic
age arguments --- see e.g. \citet{pc03} for a review --- whereas for larger
sources, i.e. MSOs \citep{murgia99} and LSOs \citep{al87, liu92}, spectral
ages have been estimated.

The lobes of a large-scale RLAGN are huge reservoirs of energy --- the
minimum energy stored in the lobes is $\sim$$10^{60}$ to $10^{64}$\,ergs
\citep[see e.g.][]{rich98} --- so even if the energy supply from the
central engine to the hotspots and the lobes eventually cuts off, the
radio source should still be observable for a substantial period of time.
As the source gradually fades, its spectrum becomes steeper and steeper
because of radiation and expansion losses. This so-called `coasting phase'
of the lobes of a RLAGN, which can last up to $10^{8}$~years \citep{kom94,
sl01}, provides information on past nuclear activity.
LSOs possessing these features are sometimes termed `faders' and,
although relatively rare,
have been observed, mostly in surveys of ultra-steep spectrum
sources \citep{rot94, db00}. B2\,0924+30 is a good 
example of this and \citet{cor87}, in describing the structure and
properties of this double radio source as a possible relic radio galaxy with
a projected linear size of $270 h^{-1}$\,kpc, has
indicated that the source is a `dying' LSO object. Recently, \citet{jam04} have
confirmed that B2\,0924+30 is in fact a relic radio structure associated
with an E/S0 galaxy, IC\,2476, that `switched off' its activity
$\sim$$5\times10^7$ years ago and as such can be labelled a `dead' radio
galaxy. The evolving spectra of the radio relics believed to be remnants of
powerful radio galaxies have been studied by \citet{gr94} and \citet{kc02}.

A question that naturally arises is whether the activity periods of
galaxies can be much shorter than those pertinent to the LSOs.
Intuitively, the answer to this question might be positive and
\citet{rb97} --- hereafter RB97 --- have proposed a model in which
extragalactic radio sources are intermittent on timescales of $\sim$$10^4$
-- $10^5$~years. According to RB97, when the power supply cuts off, the
radio emission fades rapidly. However, the shocked matter continues to
expand supersonically and keeps the basic source structure intact. Thus,
this model predicts that there should be a large number of MSOs that are
weaker than those currently known, because of this power cut-off, 
and which should have steep spectra with no sign of active nuclei.
If medium-scale, i.e. subgalactic-scale, faders could be
found, then the theory developed by RB97 would be strongly supported.
Looking for `weak' MSOs might also lead to the establishment of a more
complete evolutionary scheme for radio sources in which a number of them do
not follow the whole evolutionary track proposed by \citet{r96} but leave it
at the MSO stage.

A search for medium-sized faders was one of the reasons for the
establishment and observation of a sample of weak CSS sources using,
as a basis, the early results from the {\it Faint Images of Radio Sky at
Twenty cm} (FIRST) survey \citep{wbhg97}\footnote{Official website:
http://sundog.stsci.edu}. A number of possible MSOs were identified and in
this paper, the second in a series covering observations of the initial
sample, the results of MERLIN and VLA observations of those
sources considered to be weak MSOs are presented.

\begin{figure*}
\centering
\includegraphics[width=8cm, height=11cm]{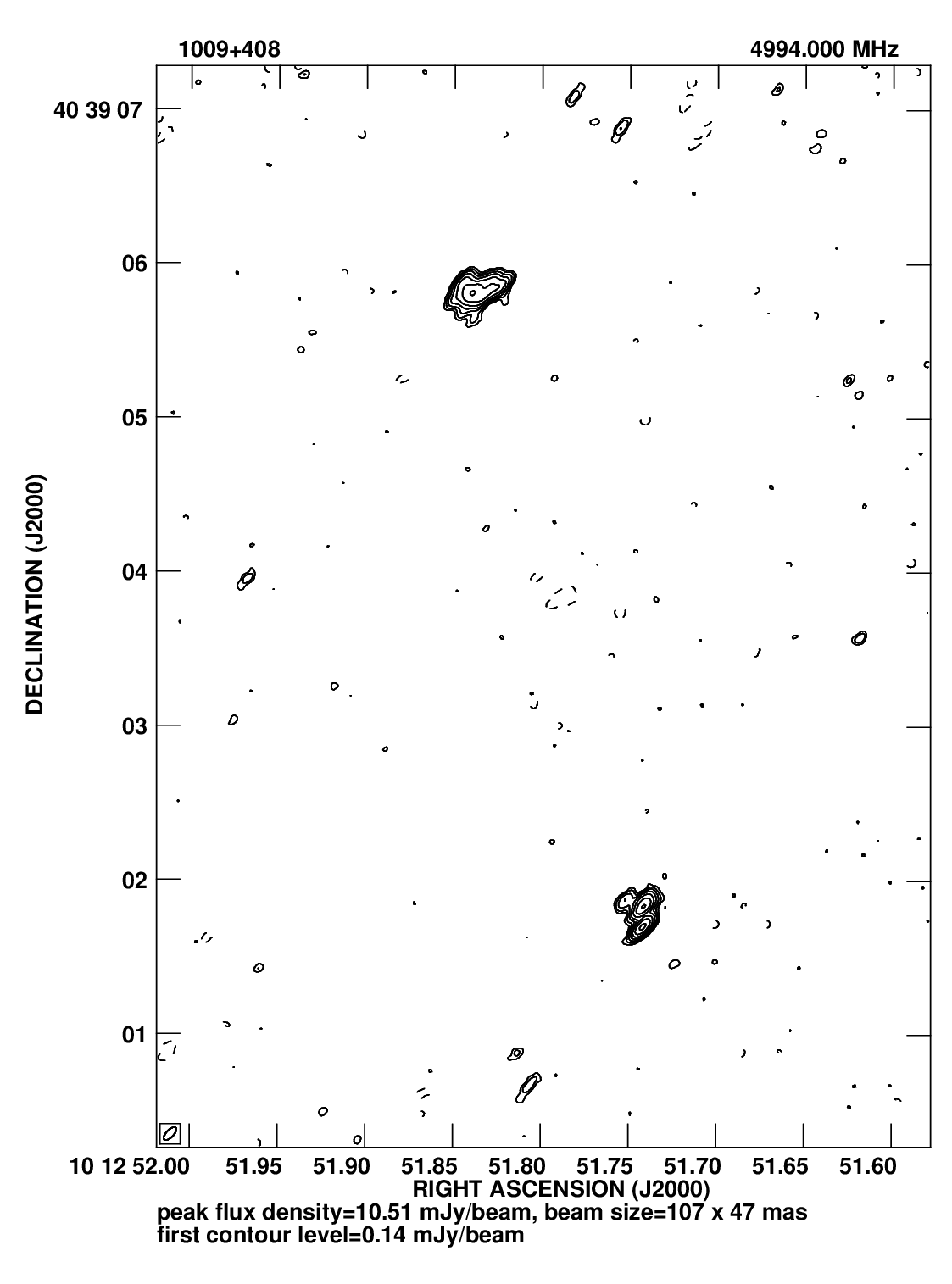}
\includegraphics[width=8cm, height=11cm]{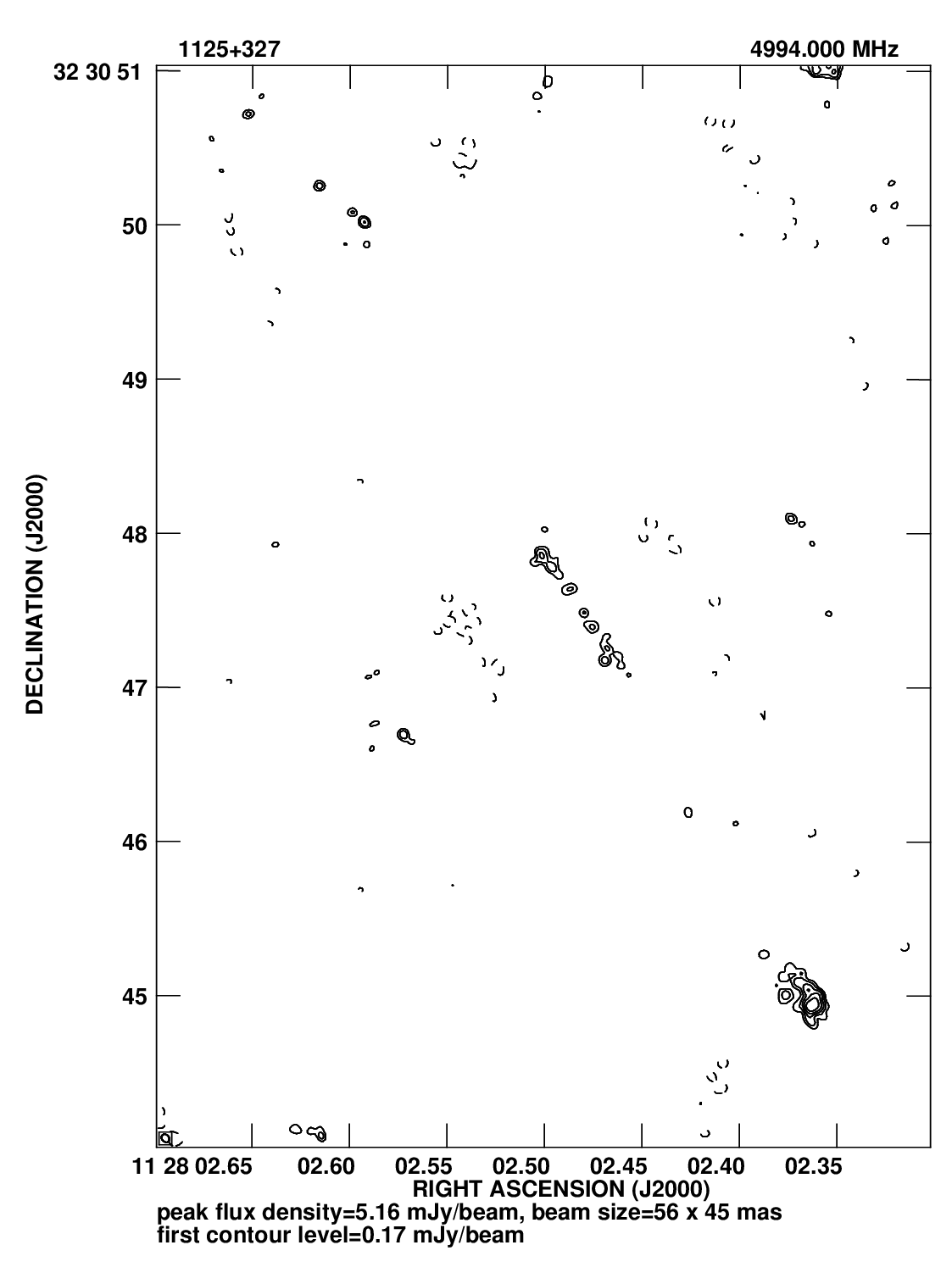}
\includegraphics[width=8cm, height=8.5cm]{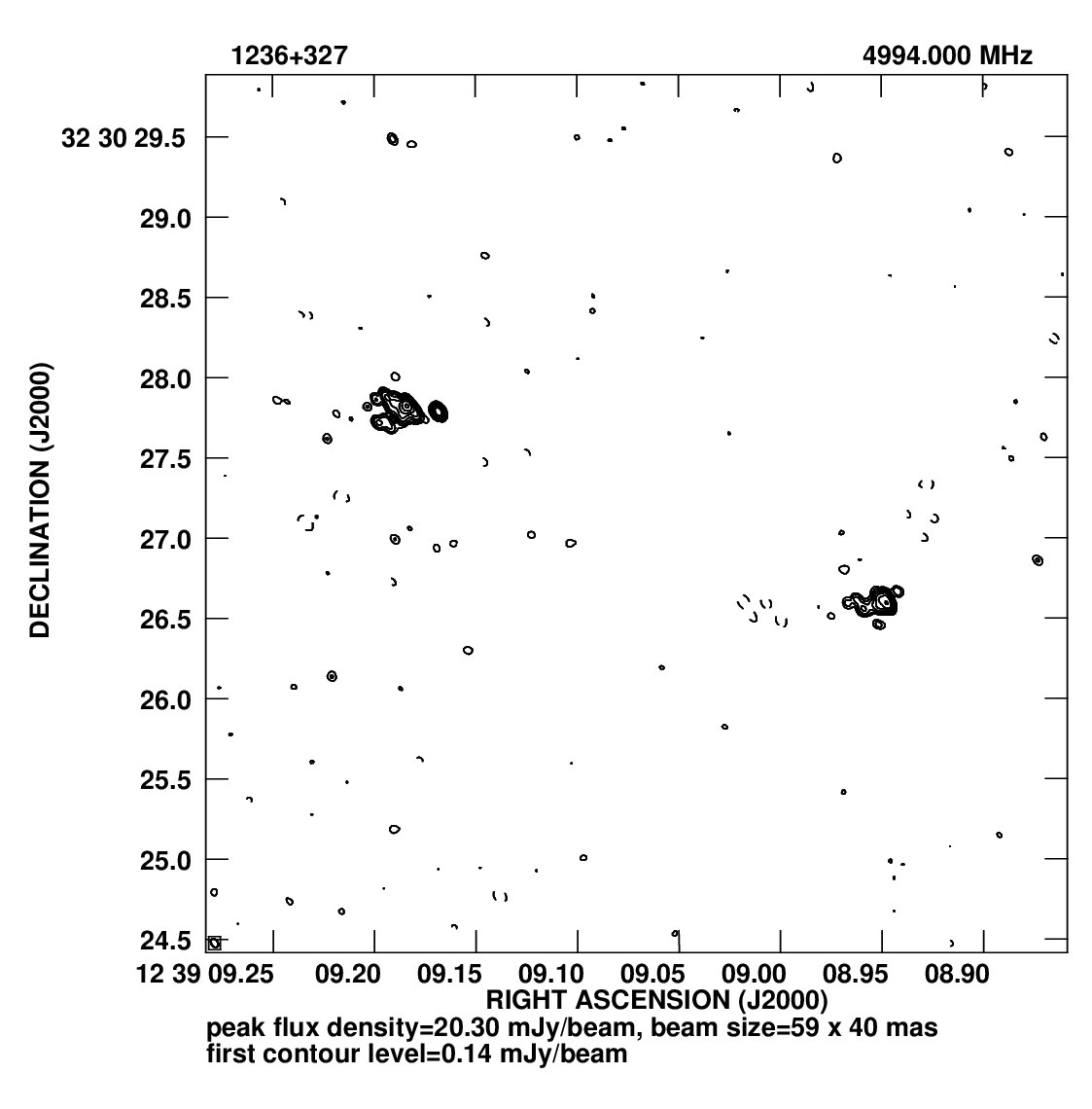}
\includegraphics[width=8cm, height=8.5cm]{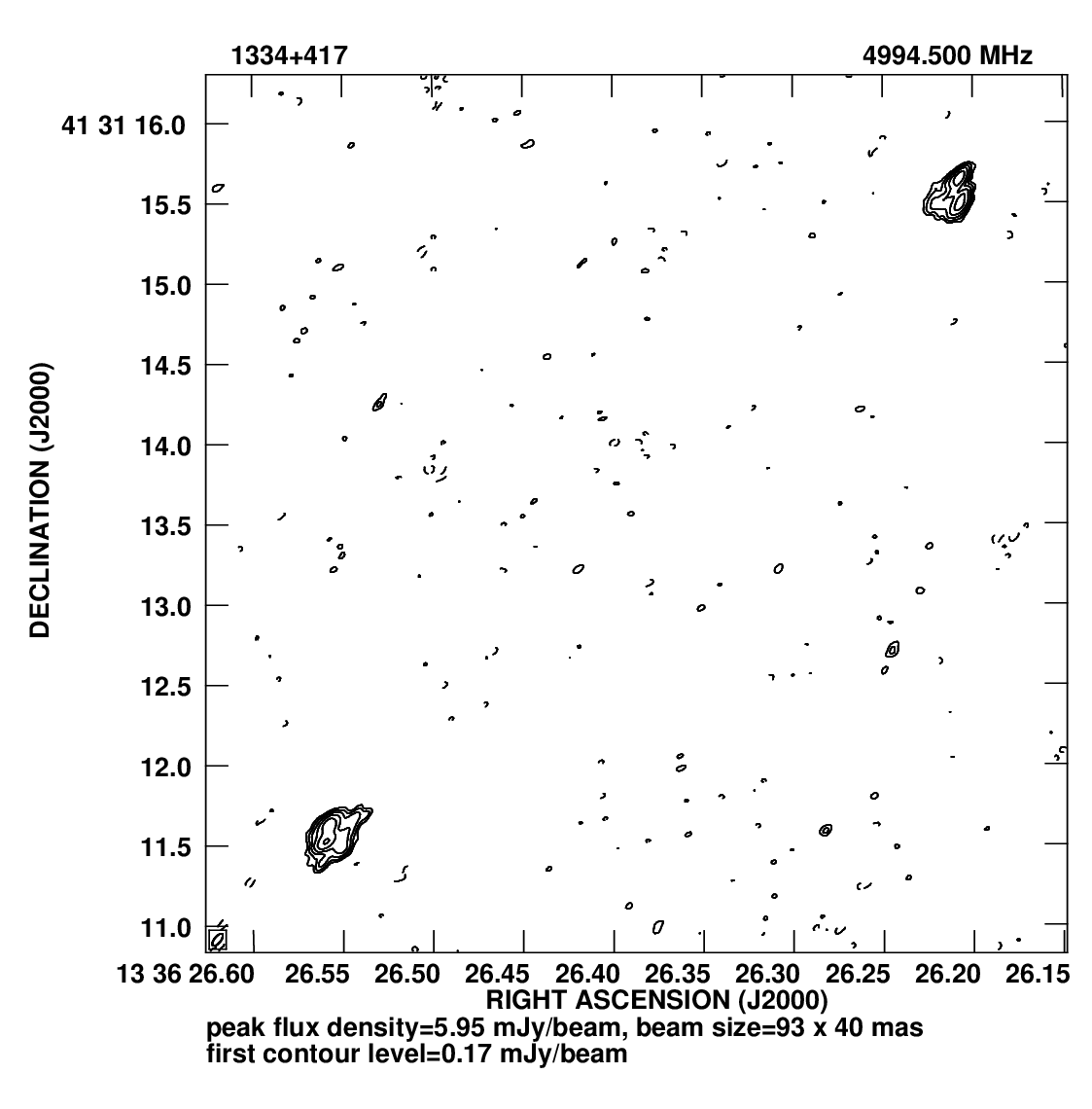}
\caption{MERLIN maps at 5\,GHz resulting from `snapshot' observations
(the `pilot' survey). Contours increase by a factor 2 and the
first contour level corresponds to $\approx 3\sigma$.}
\label{MERLIN_maps}
\end{figure*}

\begin{figure*}
\centering
\includegraphics[width=7cm, height=11.5cm]{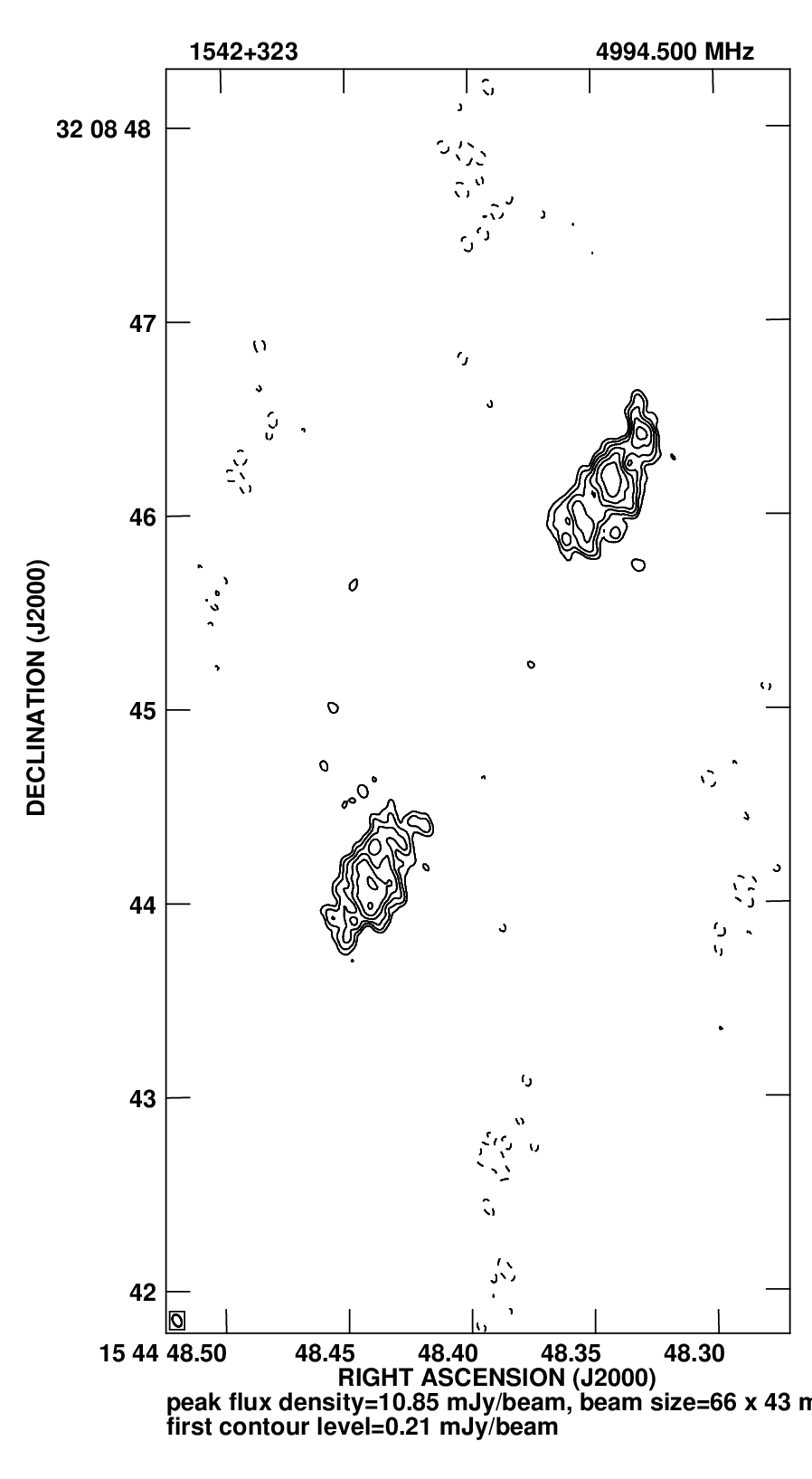}
\includegraphics[width=7cm, height=11.5cm]{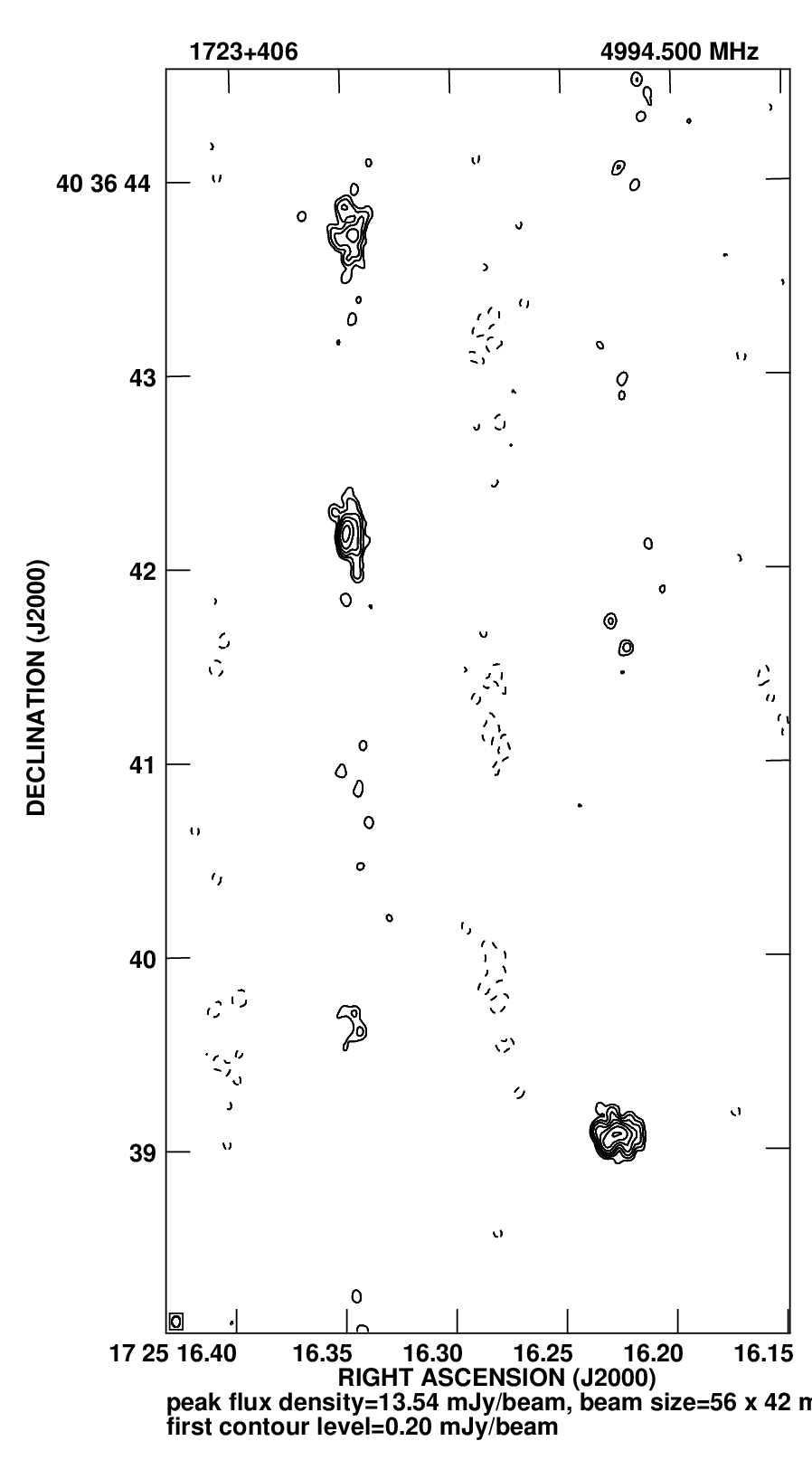}
{\small\\
{\bf Fig.\,\thefigure{}.}~(continued)}
\end{figure*}

\section{The parent sample and its previous observations}

During the period following the publication of the very first paper defining
a class of CSS sources \citep{pw82} up to the year 2000, only two almost
identical surveys of CSS sources \citep{spencer89,fanti90}, which were based
upon the 3CR catalogue \citep{ben62} and
the \citet{pw82} list, have been used in investigations of the whole class.
More recently a major step towards weaker sources has been made by
\citet{tsch03} and earlier by \citet{fanti01} who carried out VLA
observations of candidate sources selected from the B3-VLA \citep{vig89} survey.
It became clear to us that the FIRST
survey, because of its resolution (5\farcs4) and its low sensitivity limit,
could contain many weak CSS sources. Using an early release of FIRST covering
a strip of sky defined by right ascension in the range $7^h30^m <\alpha
< 17^h30^m$ and declination in the range $28\degr <\delta < 42\degr$
we selected a flux density limited, complete
sample consisting of $\sim$60~candidate sources fulfilling the basic criteria
of the CSS class: steep spectra between $\sim$0.4\,GHz and 5\,GHz and high
compactness. A detailed description of the selection criteria was
given in \citet{kun02} --- hereafter Paper~I. The candidates were, of
course, much weaker than other known objects selected in a similar manner;
their 5-GHz fluxes listed in the GB6 survey \citep{bec91} were in the range 
150\,mJy $< S_{5\,GHz}<$ 550\,mJy. Our sample is therefore at a comparable 
depth to that of \citet{fanti01}.

The sample was initially observed with MERLIN in `snapshot' mode at 5\,GHz
(a `pilot' survey), the details of which, including typical $u$-$v$
coverage for the sources, were given in Paper~I. Final images of five
arcsecond-scale sources were also presented in Paper~I, only one of
which was a MSO with conspicuous hotsposts embedded
within its FR\,II-like structure. Many of the images
acquired in the pilot survey showed the need for follow-up
observations of selected groups of objects using, appropriately, the
VLA, MERLIN, the EVN and the VLBA. Six sources were identified as MSOs and
potential candidates for scaled-down versions of faders. They appeared to be
unaffected by beaming and Doppler boosting and had simple, double
structures typical for MSOs, yet without clear indications of hotspots.
The MERLIN 5-GHz `snapshot' images of these six
MSOs are presented in Fig.~\ref{MERLIN_maps}.
The details and results of further observations of these six sources with
MERLIN and the VLA are also presented in this paper (Paper~II). Observations
of 19~other sources with angular sizes below 1\arcsec, many of them made using
VLBI techniques, will be presented in a forthcoming paper (Paper~III).

\section{Follow-up observations and their reduction}

It was realised that the 5-GHz MERLIN snapshot
images of the pilot survey might have suffered from an undersampling of 
the $u$-$v$
plane and therefore some flux could be missing. To overcome this problem,
the sources were observed with the VLA in A-configuration at 4.9\,GHz.
MERLIN 1.7-GHz observations were also made in an attempt to
determine the spectral indices of the lobes.

The VLA observations at 4.9\,GHz ($0\farcs4$ resolution) were carried
out on 29 and 30 June 2003 in `snapshot' mode. Each target source was 
observed four times in
4-minute scans interleaved with 1-minute scans on phase calibrators. 3C286
was used as a primary flux density calibrator and the whole data reduction
process was carried out in AIPS using the standard phase and amplitude
calibration tasks. The task SCMAP, initialised for a few passes of phase
self-calibration, was used to produce `naturally weighted' images.
The total intensity images are shown in Figs.~2a-f (left panels).

The 1.7-GHz MERLIN observations (resolution $\sim$0\farcs15) 
were carried out in February and March 2003. Each target source scan was
interleaved with a scan on a phase reference source throughout
an $\sim$14-h track, except for the two sources 1236+327 and 1334+417. The
total cycle time (target -- phase-reference) was 6 minutes including
telescope
drive times, with $\sim$3.5\,minutes actually on the target source per cycle.
The increased sensitivity of MERLIN resulting from the inclusion of the
Lovell telescope in the observations of 1236+327 and 1334+417 meant that
both
these sources and their associated phase-references could be interleaved
at 1-hour intervals in a single $\sim$14-hour track. The phase-reference 
sources
were chosen from the MERLIN calibrator lists derived from the JVAS survey
\citep{jvas1, jvas2, jvas3}. OQ208 was used as the point source or
baseline calibrator and 3C286 as the flux and polarisation calibrator.
The flux density for 3C286 at 1658\,MHz on the VLA scale, which was used,
was 13.639\,Jy.

The initial editing and both amplitude and polarisation
calibration of the data were carried out using the
Jodrell Bank d-programmes and PIPELINE automated procedure.
Further cycles of
phase self-calibration and imaging using the AIPS tasks CALIB and IMAGR
were then used to produce the final total power and, where available,
the polarisation images shown in Figs.~2a-f (right panels).

The snapshot images of \object{1236+327} and \object{1542+323}
showed that both sources deserved further investigation.
Consequently, MERLIN full-track, phase-referenced observations of them were
carried out on 18 and 24 May 2004 at 5\,GHz and 4.9\,GHz respectively.
The high fidelity images resulting from these observations, as well as
combined MERLIN+VLA images, are shown in 
Figs.~\ref{1236+327_MERLIN_5GHz.map} and \ref{1542+323_MERLIN_5GHz.map}.

The rms noise levels in the MERLIN 1.7-GHz images are significantly greater
than in the VLA 4.9-GHz ones. The shortest MERLIN $u$--$v$ spacing is
also only 12\,km, which limits its sensitivity to more extended emission.
Consequently, for both of the above reasons, attempts to produce
detailed spectral index images from the MERLIN 1.7-GHz and VLA 4.9-GHz
images, particularly in the regions of low-level extended emission, were
unsatisfactory and the results have not been included.

\begin{table*}[t] 
\caption[]{Optical magnitudes derived form POSS plates using the APM and
radio flux densities of 6~CSS sources at 1.4 and 4.85\,GHz}
\begin{center}
\begin{tabular}{@{}c c c l c c c c c c@{}}
\hline
\hline
       &    &     &   &   &Total & Total&   &Total  &  \\
~~~Source & R.A. & Dec. & ID&$m_{R}$&flux at & flux at&
$\alpha_{1.4\mathrm{GHz}}^{4.85\mathrm{GHz}}$& flux at &LAS~~~ \\
~~~Name   & h~m~s & $\degr$~$\arcmin$~$\arcsec$ &  & &1.4 GHz &4.85 GHz&
&1.7 GHz &$\arcsec$~~~\\
       &     &   &     &   & mJy&mJy& &mJy& \\
~~~(1)& (2)& (3) &(4)& (5) &(6)& (7)& (8) &(9)&(10)~~~\\
\hline
\hline
~~~1009+408& 10 12 51.795& 40 39 03.73&G
&20.40&411&194&$-$0.60&366&4.06~~~\\

~~~1125+327& 11 28 02.464& 32 30 46.70&G&19.11
&600&205&$-$0.86&508&6.81~~~\\

~~~1236+327& 12 39 09.090& 32 30
27.45&EF&---&832&256&$-$0.95&692&3.12~~~\\

~~~1334+417& 13 36 26.394& 41 31
13.37&G&22.62&470&154&$-$0.90&395&5.47~~~\\

~~~1542+323& 15 44 48.395& 32 08
45.11&G&21.14&854&325&$-$0.78&734&2.41~~~\\

~~~1723+406& 17 25 16.268& 40 36
41.38&EF&---&962&221&$-$1.18&765&4.65~~~\\
\hline
\end{tabular}
\end{center}

\vspace{0.5cm}
{\small
Description of the columns:
\begin{itemize}
\item[---]{Col.~~(1): Source name in the IAU format;}
\item[---]{Col.~~(2): Source right ascension (J2000) extracted from FIRST;}
\item[---]{Col.~~(3): Source declination (J2000) extracted from FIRST;}
\item[---]{Col.~~(4): Optical identification: G-galaxy, EF-empty
field;}
\item[---]{Col.~~(5): Red magnitude;}
\item[---]{Col.~~(6): Total flux density at 1.4\,GHz extracted from FIRST;}
\item[---]{Col.~~(7): Total flux density at 4.85\,GHz extracted from GB6;}
\item[---]{Col.~~(8): Spectral index ($S\propto\nu^{\alpha}$) between
1.4 and 4.85\,GHz calculated using flux densities in columns (6) and (7);}
\item[---]{Col.~~(9): Estimated total flux density at 1.7\,GHz interpolated
from data from columns (6) and (7);}
\item[---]{Col.~(10): Largest Angular Size (LAS) in arcseconds as estimated in
the present paper.}\\
\phantom{Col.~(10):} (LAS is measured as a separation between the
outermost components' peaks.)
\end{itemize}
}
\label{table1}
\end{table*}

\stepcounter{figure}

\begin{figure*} 
\centering
\includegraphics[height=8cm,width=8cm]{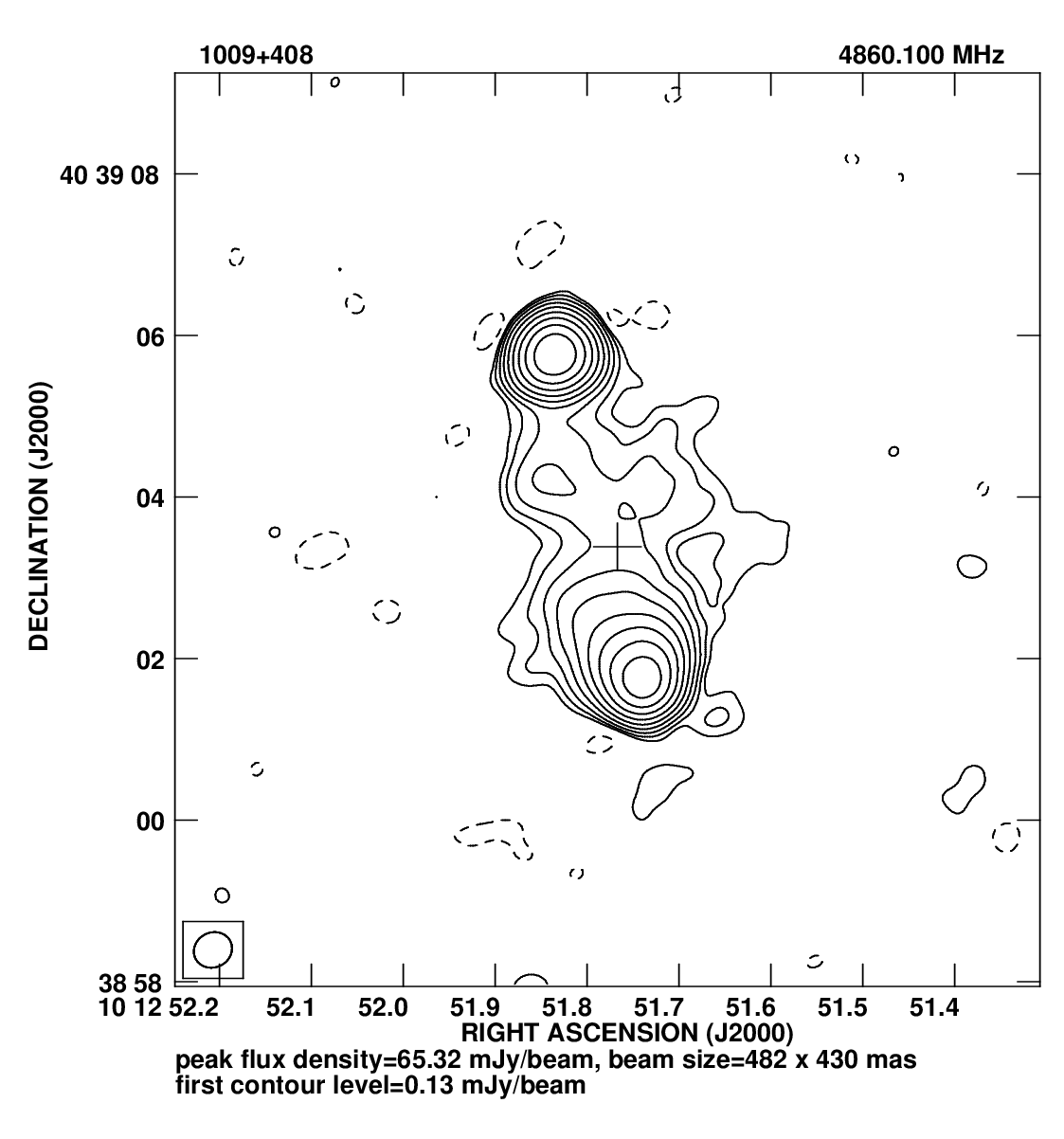}
\includegraphics[height=8cm,width=8cm]{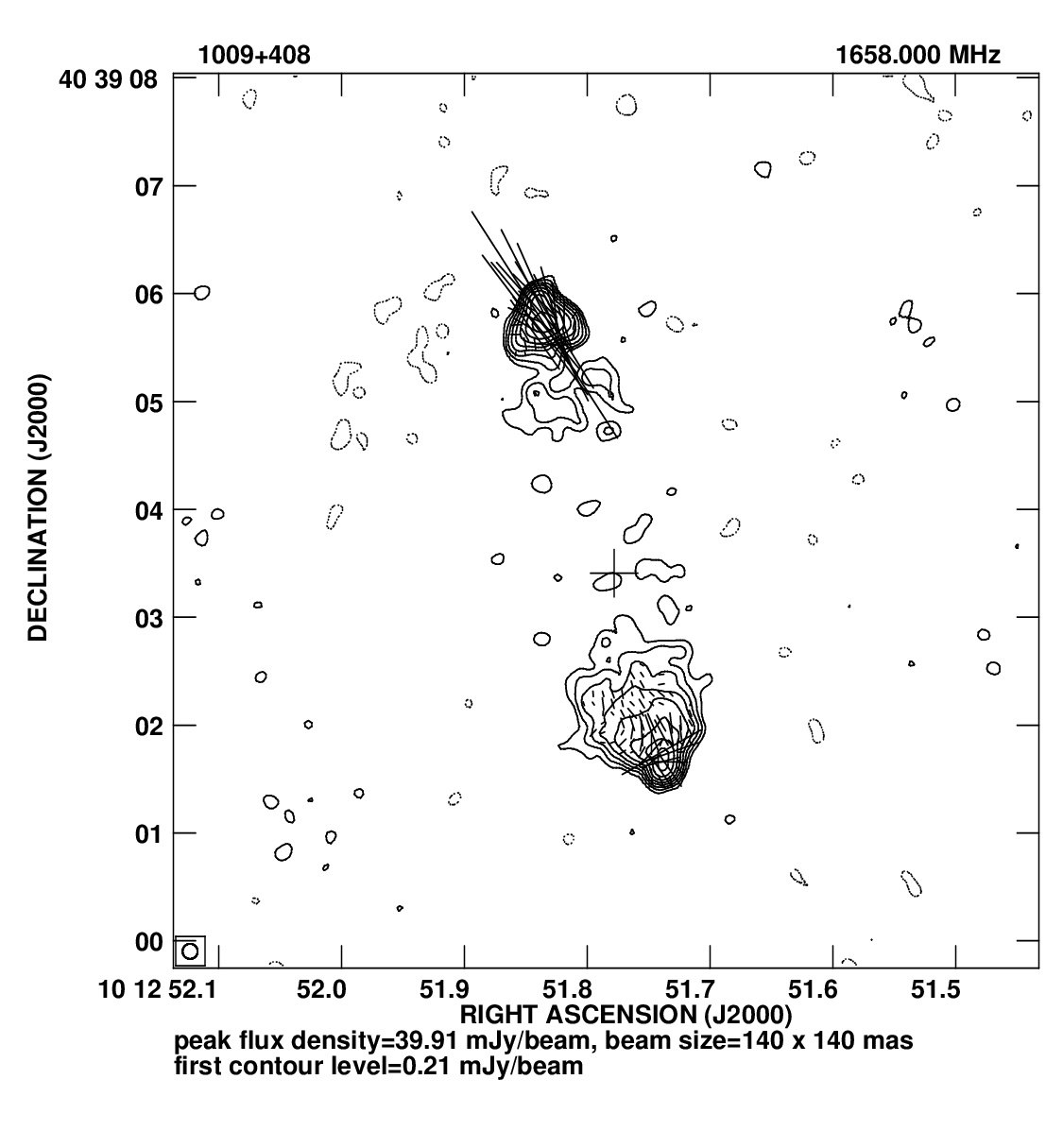}
\includegraphics[height=8cm,width=8cm]{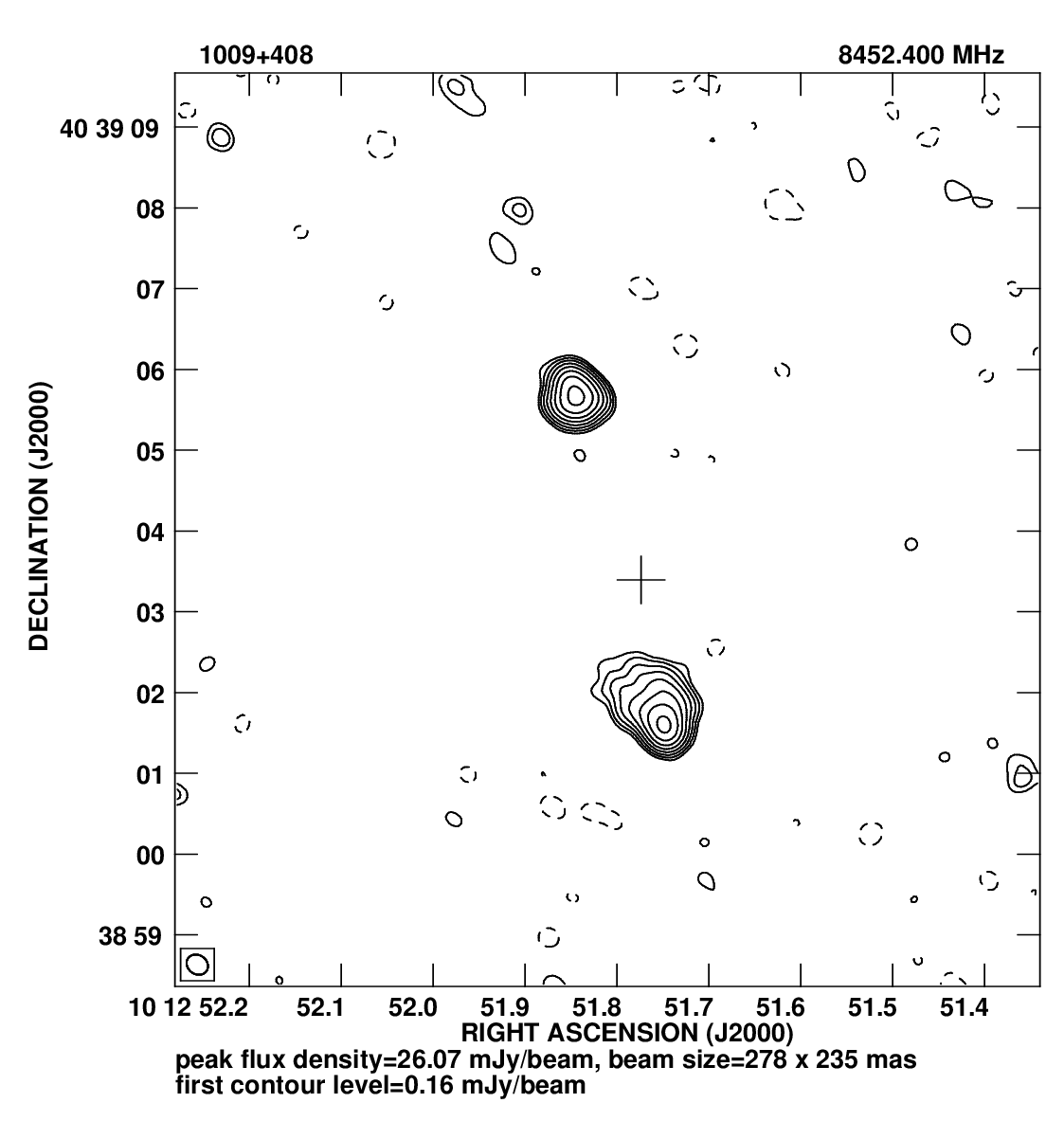}
{\small\\
Fig.\,\thefigure{}a.
VLA map of 1009+408 at 4.9\,GHz (upper left), MERLIN 1.7-GHz
polarisation map (upper right) and VLA 8.4-GHz map (lower).
Contours increase by a factor 2 and the first contour
level corresponds to $\approx 3\sigma$.
Polarisation line of $1\arcsec$ amounts to 3.3\,mJy/beam.
Crosses indicate the position of an optical object found using the SDSS/DR3.} 
\label{1009+408_maps}
\end{figure*}

\begin{figure*}
\centering
\includegraphics[height=8cm,width=8cm]{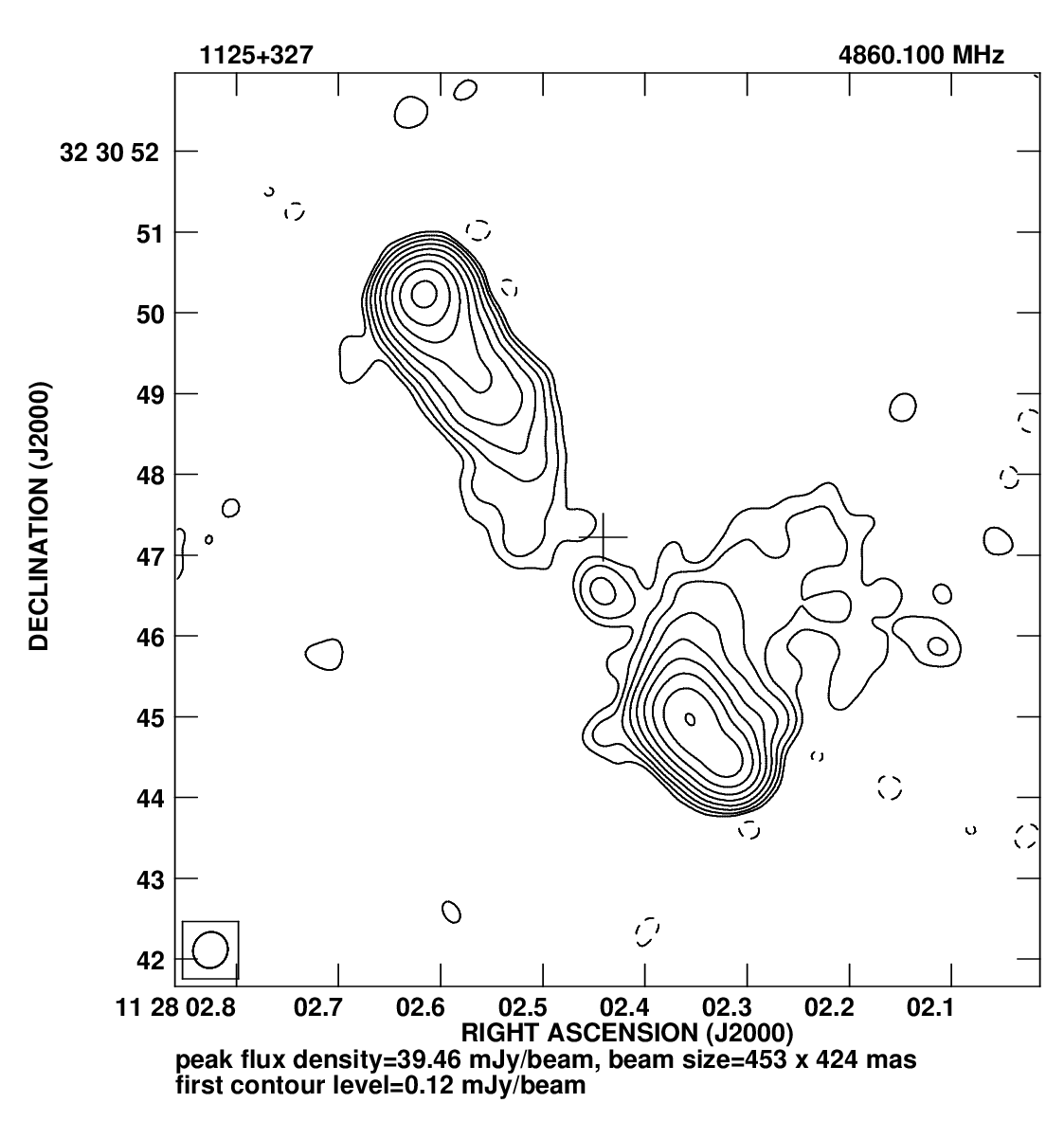}
\includegraphics[height=8cm,width=8cm]{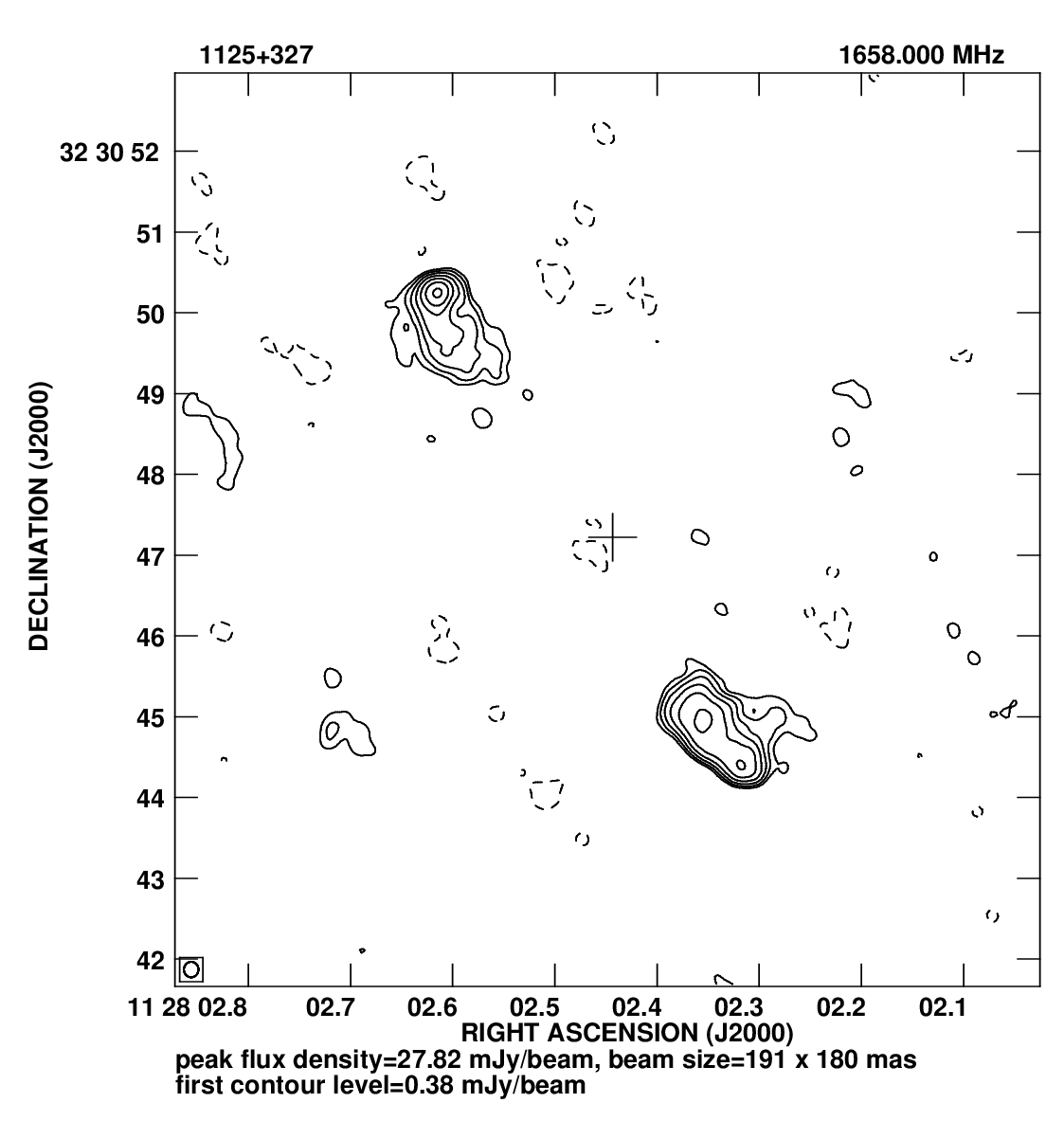}
{\small\\
Fig.\,\thefigure{}b.
VLA map of 1125+327 (left) at 4.9\,GHz and MERLIN map at
1.7\,GHz (right). Contours increase by a factor 2 and the first
contour level corresponds to $\approx 3\sigma$.
Crosses indicate
the position of an optical object found on the POSS plates using the APM.}
\label{1125+327_maps}
\end{figure*}

\begin{figure*}
\centering
\includegraphics[height=8cm,width=8cm]{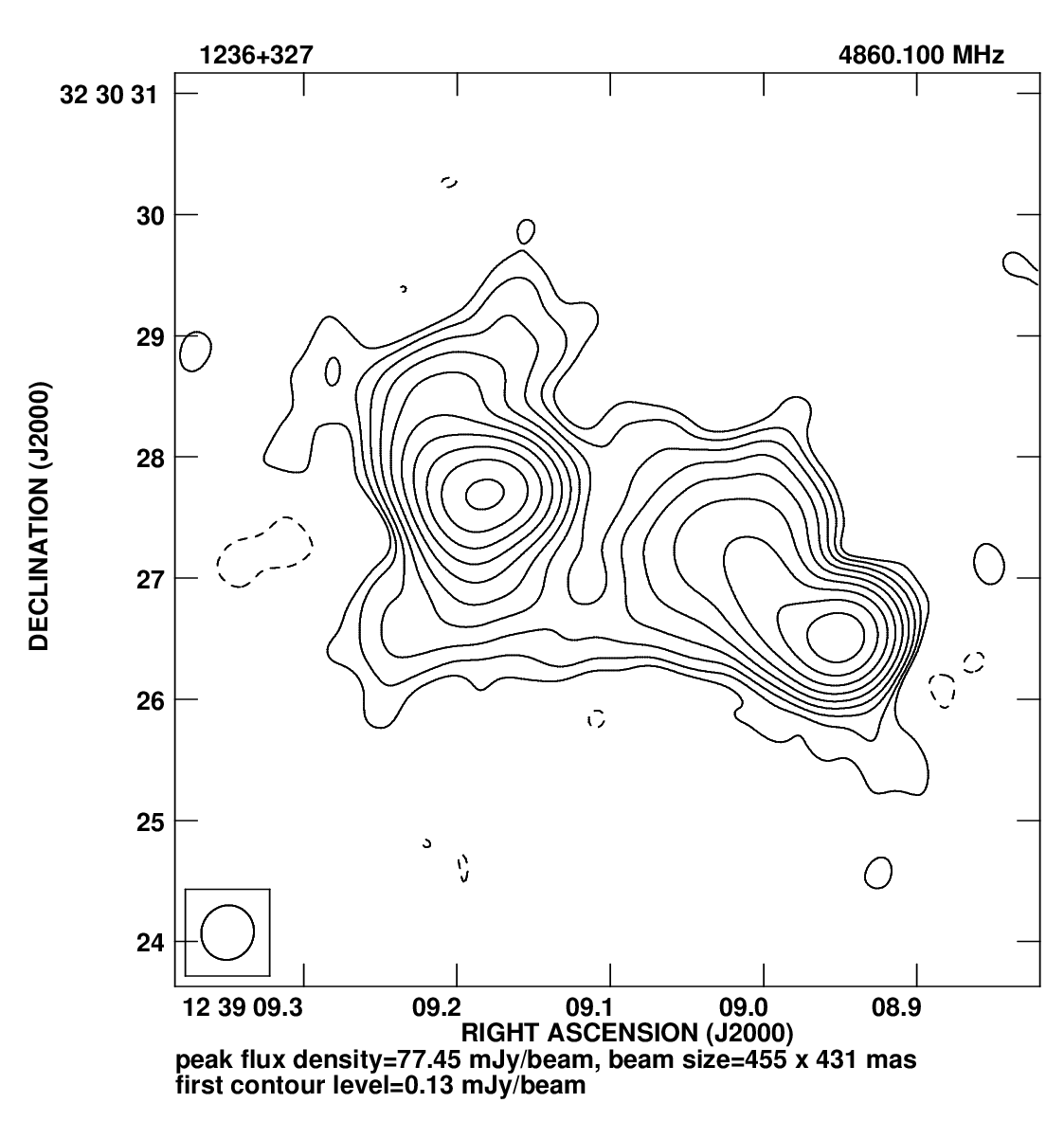}
\includegraphics[height=8cm,width=8cm]{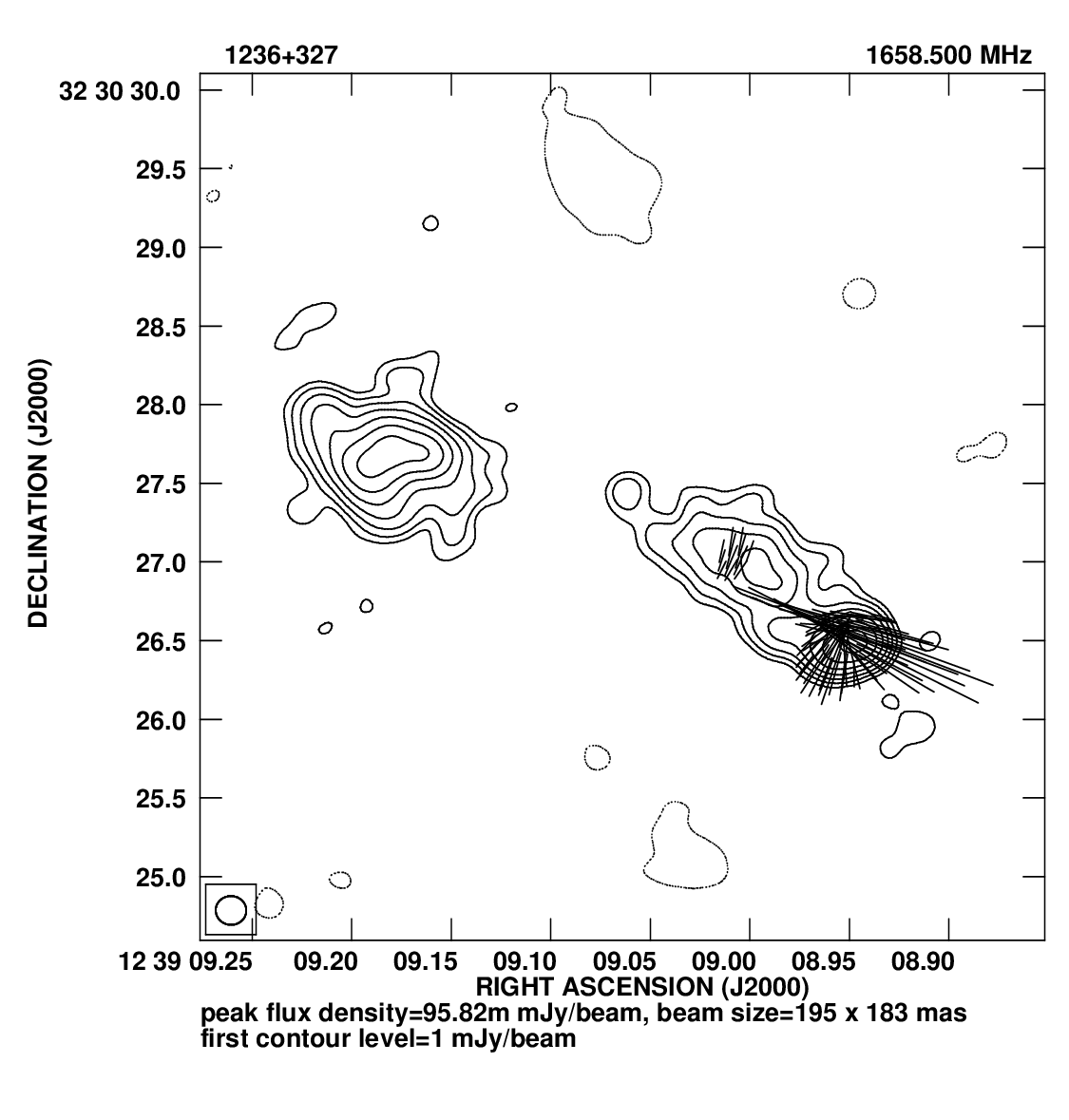}
{\small\\
Fig.\,\thefigure{}c.
VLA map of 1236+327 at 4.9\,GHz (left) and 
MERLIN polarisation map at 1.7\,GHz (right). 
Contours increase by a factor 2 and the first
contour level corresponds to $\approx 3\sigma$.
Polarisation line of $1\arcsec$ amounts
to 8.34\,mJy/beam.}
\label{1236+327_maps}
\end{figure*}

\begin{figure*}
\centering
\includegraphics[height=8cm,width=8cm]{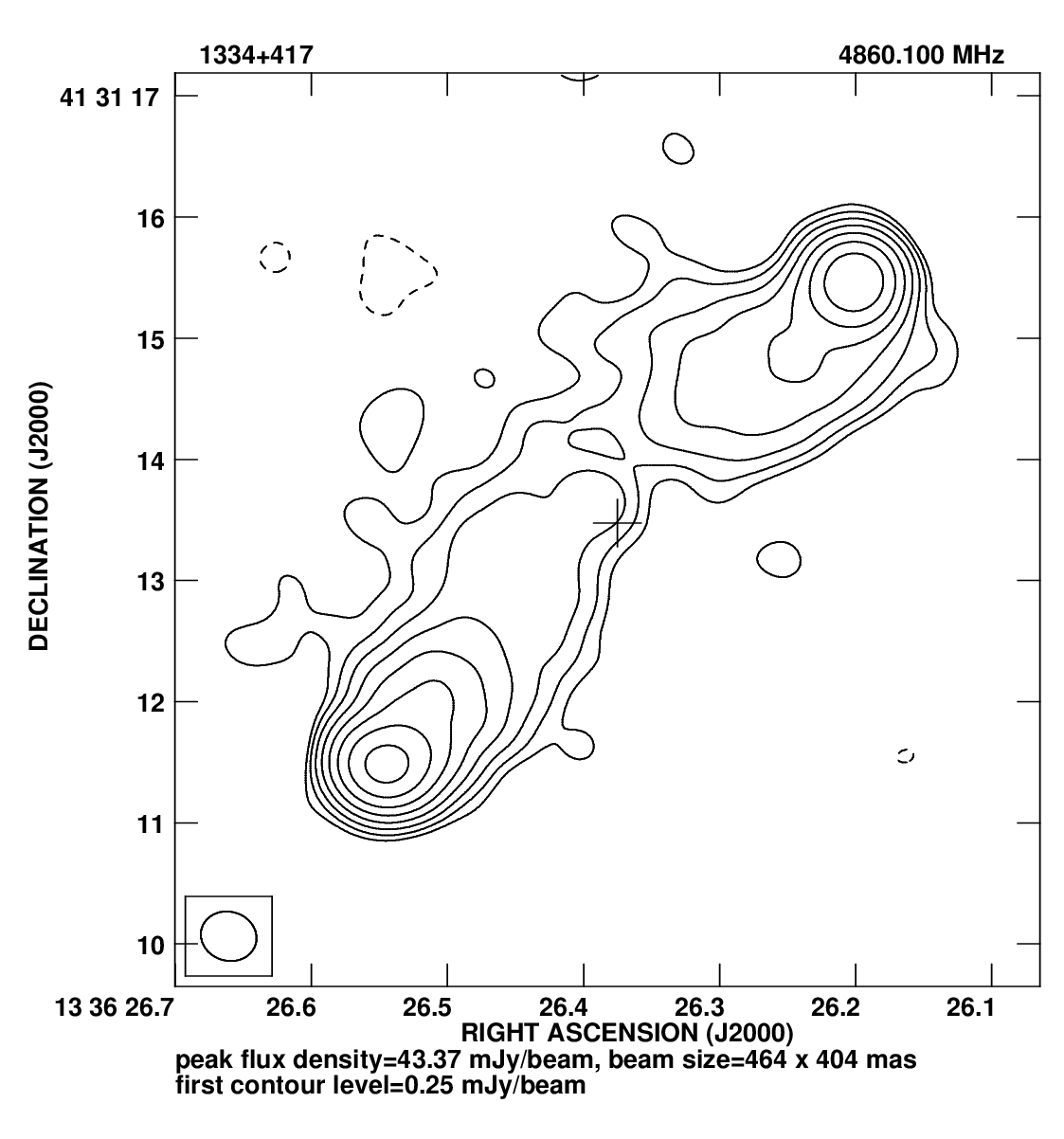}
\includegraphics[height=8cm,width=8cm]{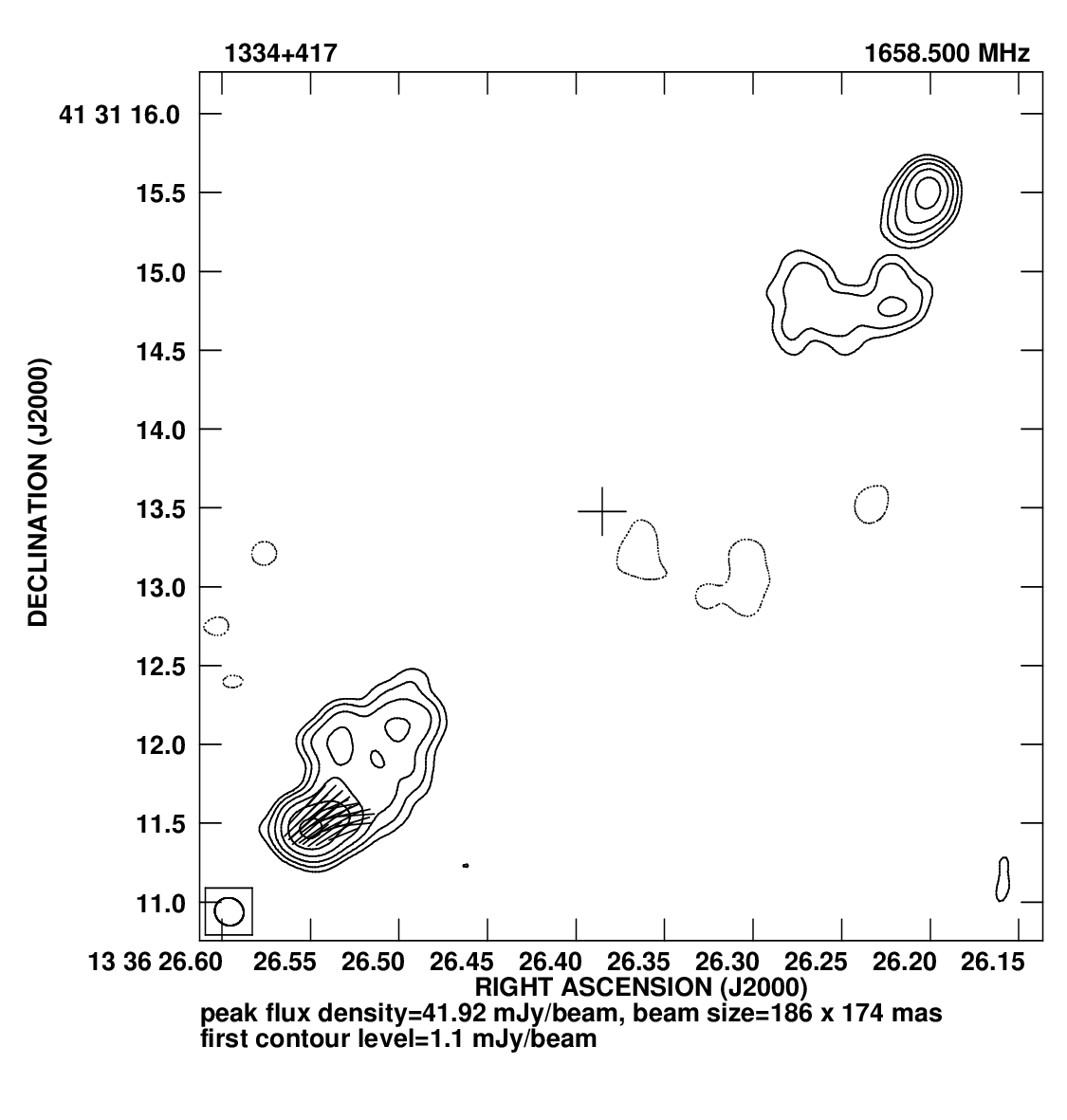}
{\small\\
Fig.\,\thefigure{}d.
VLA map of 1334+417 at 4.9\,GHz (left) and MERLIN polarisation 
map at 1.7\,GHz (right). Contours increase by a factor 2 and the first
contour level corresponds to $\approx 3\sigma$.
Polarisation line of $1\arcsec$ amounts to 6.25\,mJy/beam.
Crosses indicate the position of an optical object found using the SDSS/DR3.}
\label{1334+417_maps}
\end{figure*}

\begin{figure*}
\centering
\includegraphics[height=8cm,width=8cm]{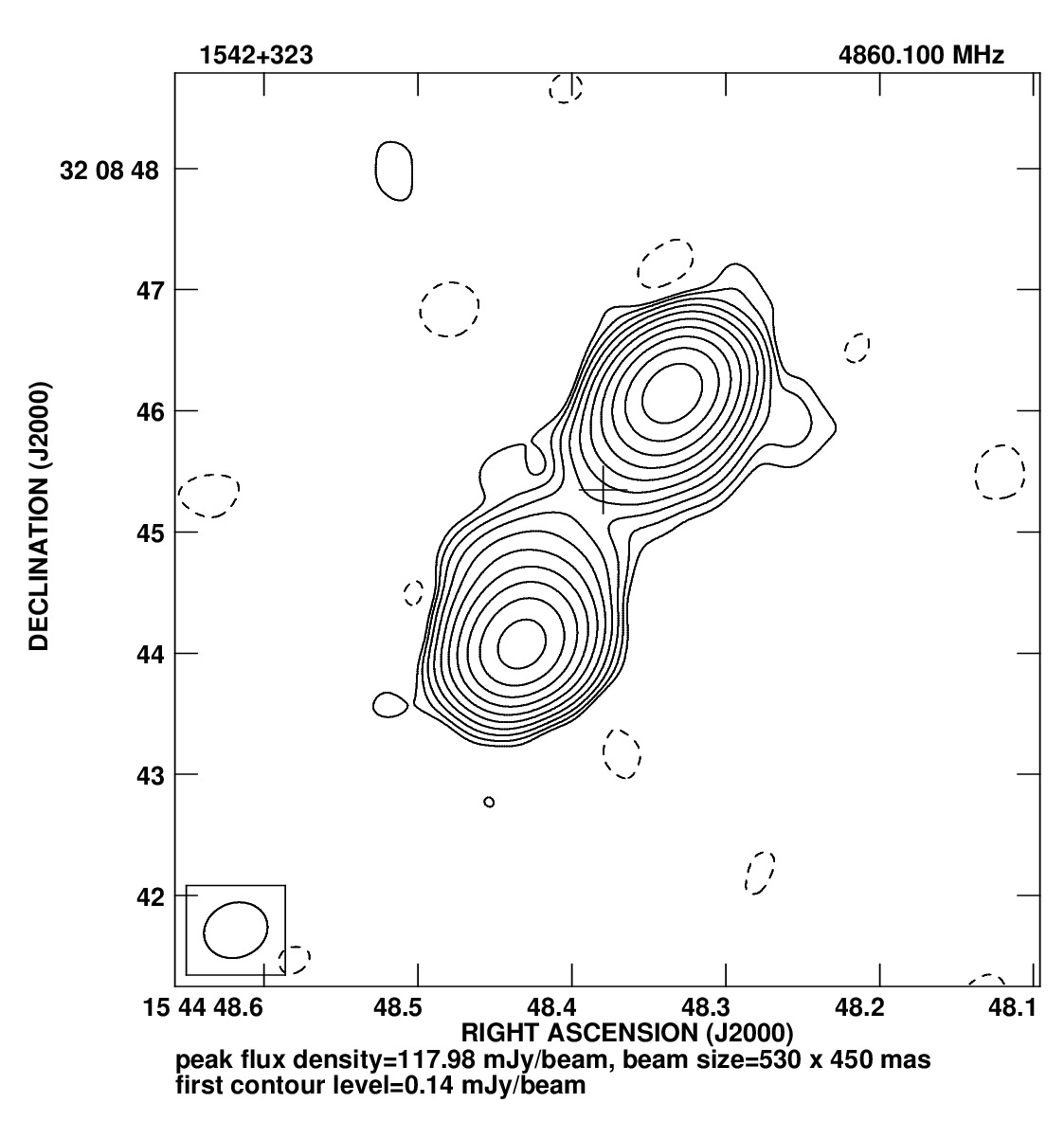}
\includegraphics[height=8cm,width=8cm]{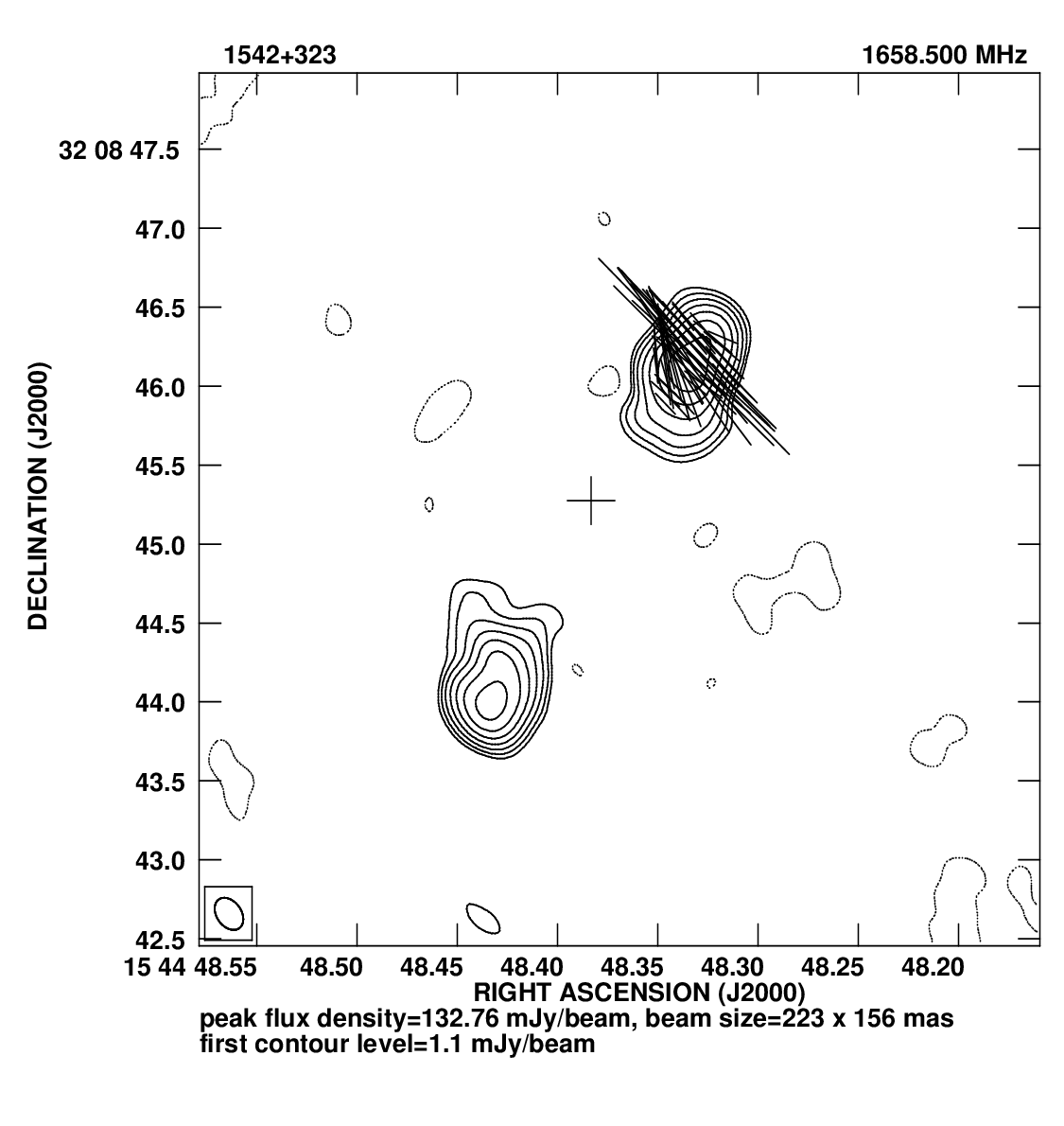}
{\small\\
Fig.\,\thefigure{}e.
VLA map of 1542+323 at 4.9\,GHz (left) and MERLIN polarisation map
at 1.7\,GHz (right). Contours increase by a factor 2 and the first
contour level corresponds to $\approx 3\sigma$.
Polarisation line of $1\arcsec$ amounts to 6.25\,mJy/beam.
Crosses indicate the position of an optical object found using the SDSS/DR3.
}
\label{1542+323_maps}
\end{figure*}

\begin{figure*}
\centering
\includegraphics[height=8cm,width=8cm]{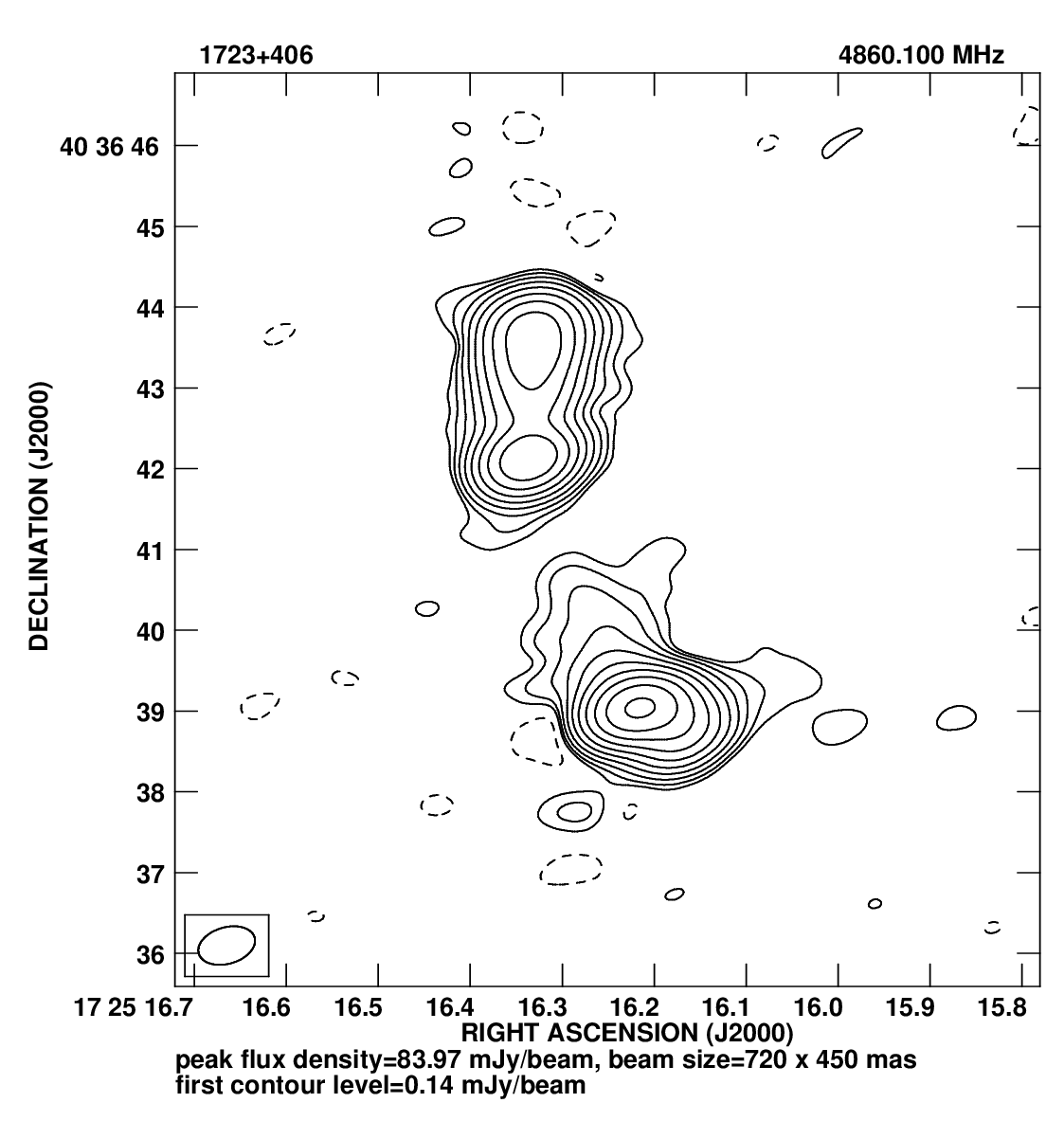}
\includegraphics[height=8cm,width=8cm]{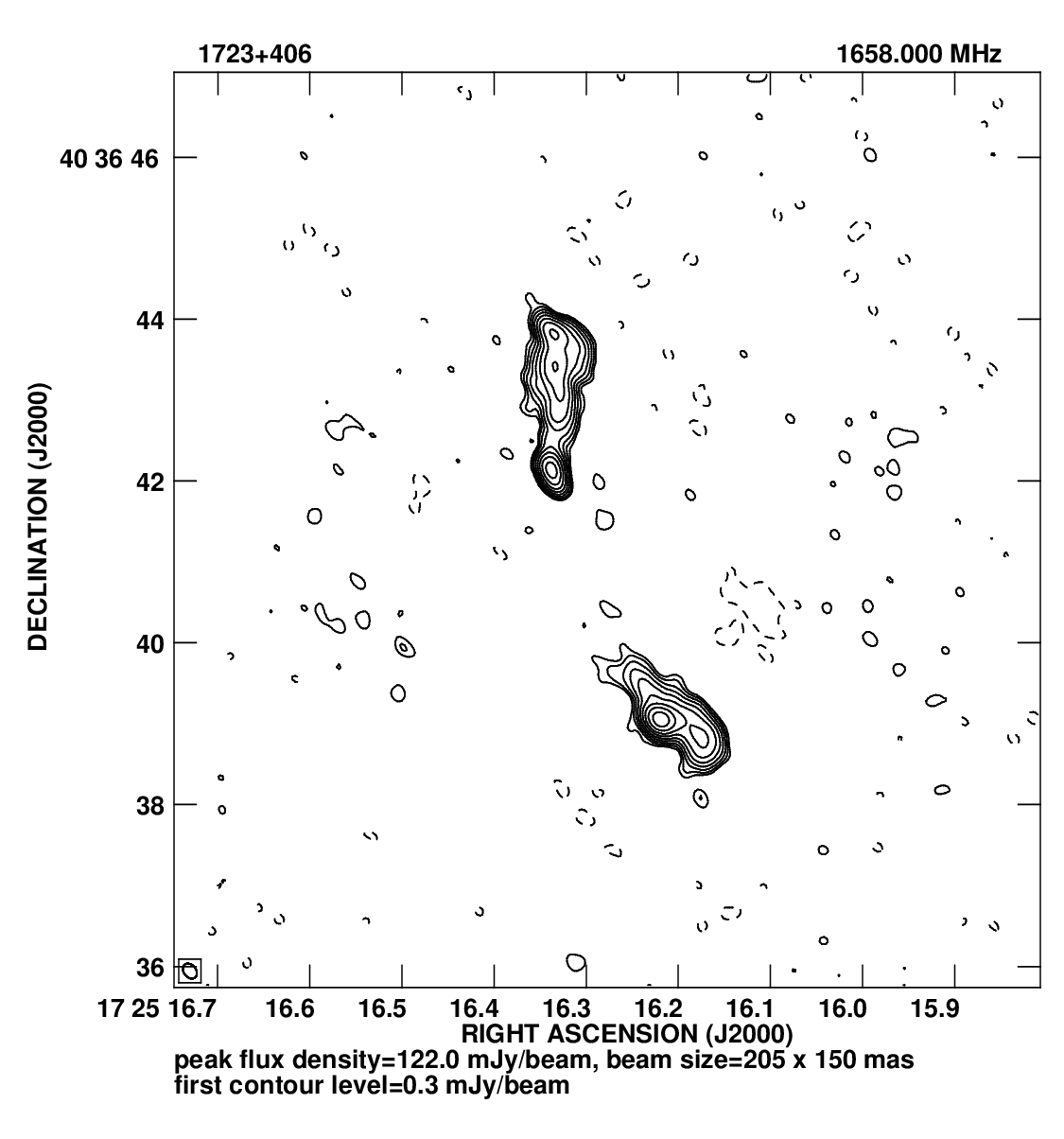}
{\small\\
Fig.\,\thefigure{}f.
VLA map of 1723+406 at 4.9\,GHz (left) and MERLIN map at
1.7\,GHz (right). Contours increase by a factor 2 and the first
contour level corresponds to $\approx 3\sigma$.}
\label{1723+406_maps}
\end{figure*}

\begin{figure*}[t]
\centering
\includegraphics[width=8cm]{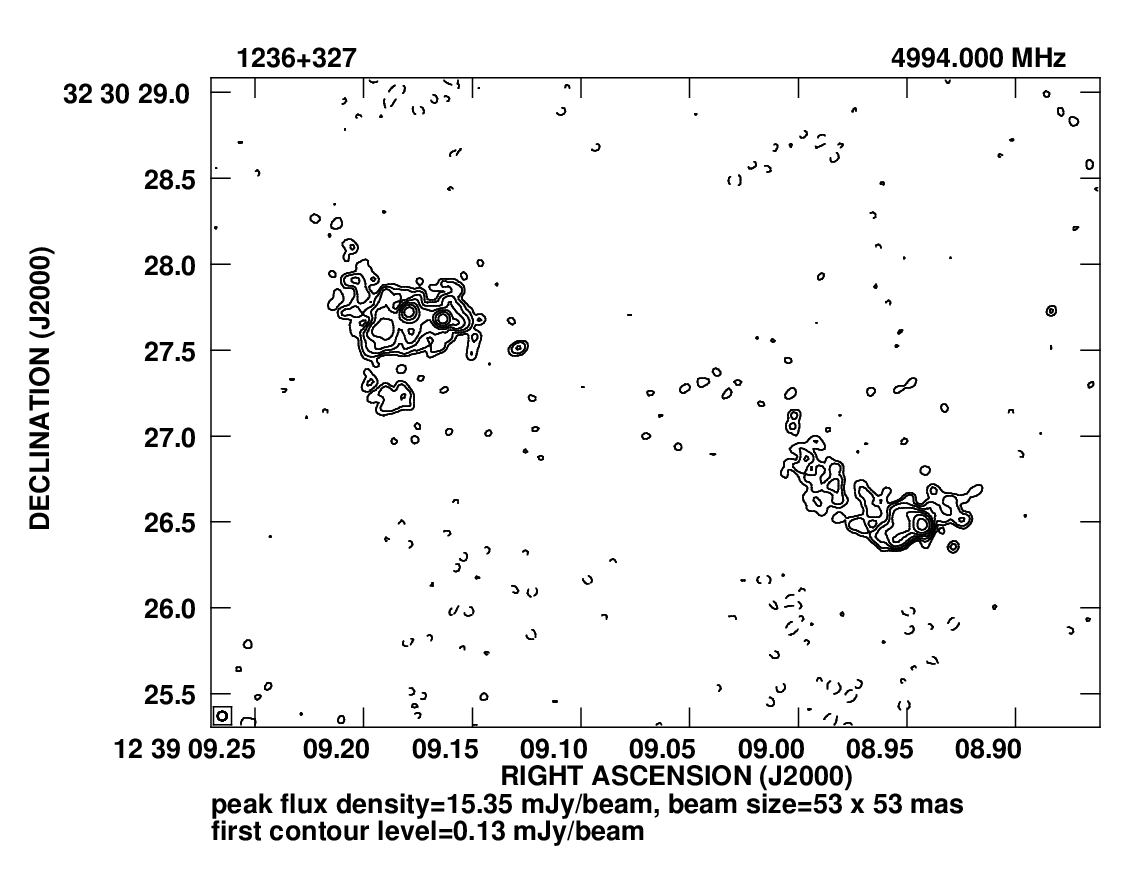}
\includegraphics[width=8cm]{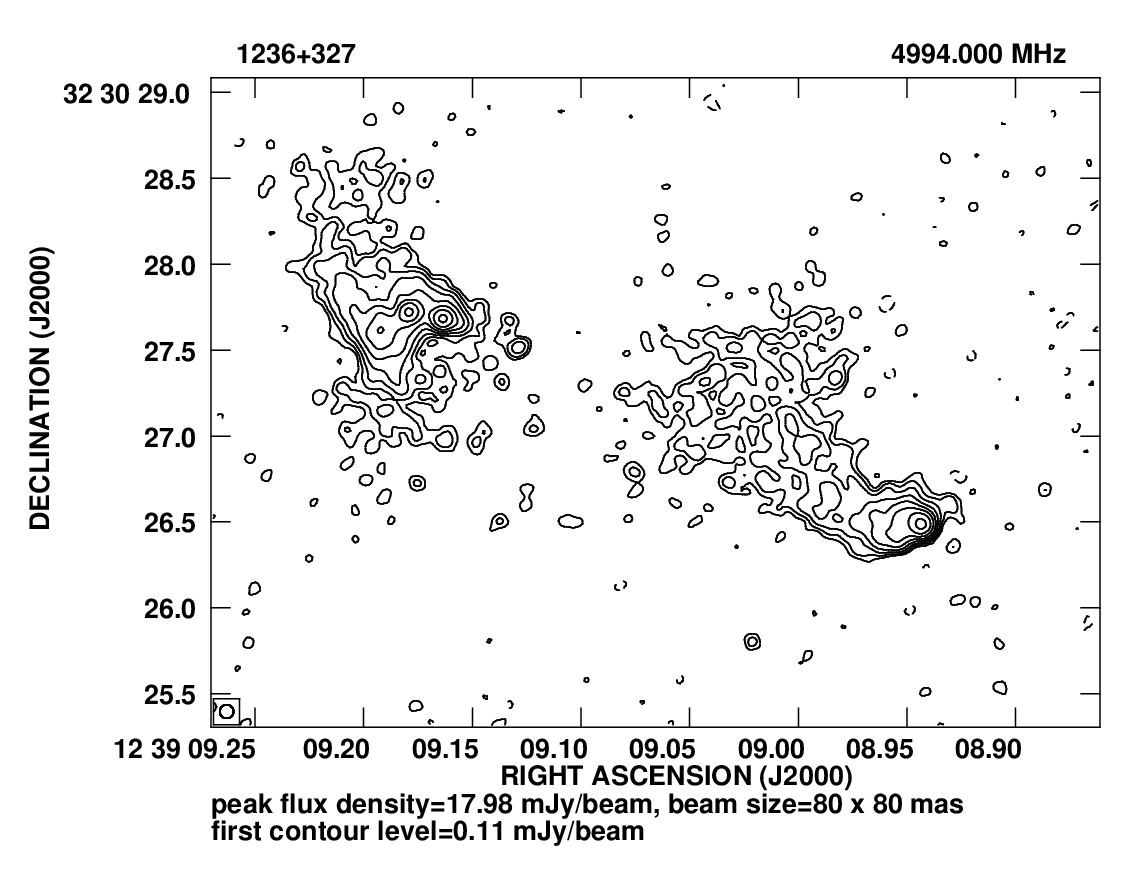}
\includegraphics[width=8cm]{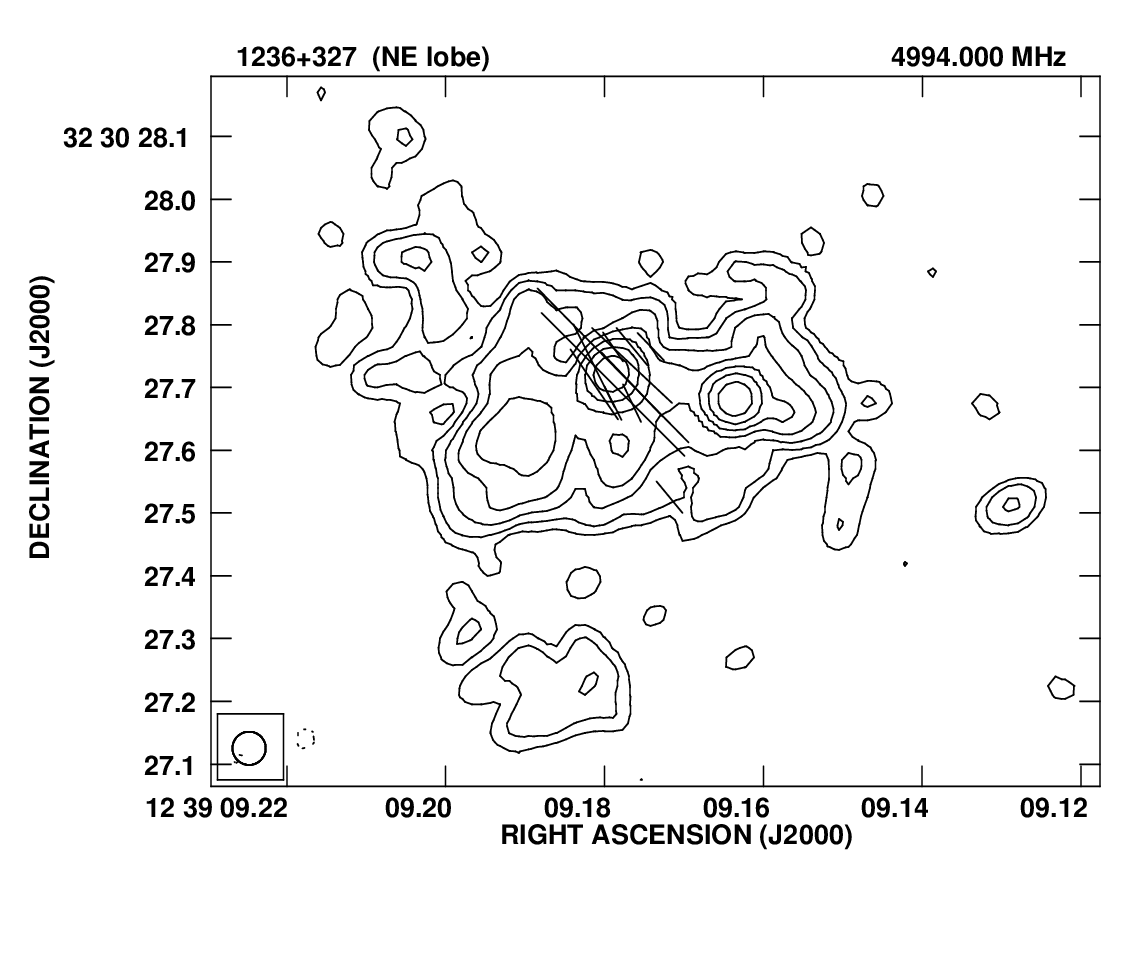}
\includegraphics[width=8cm]{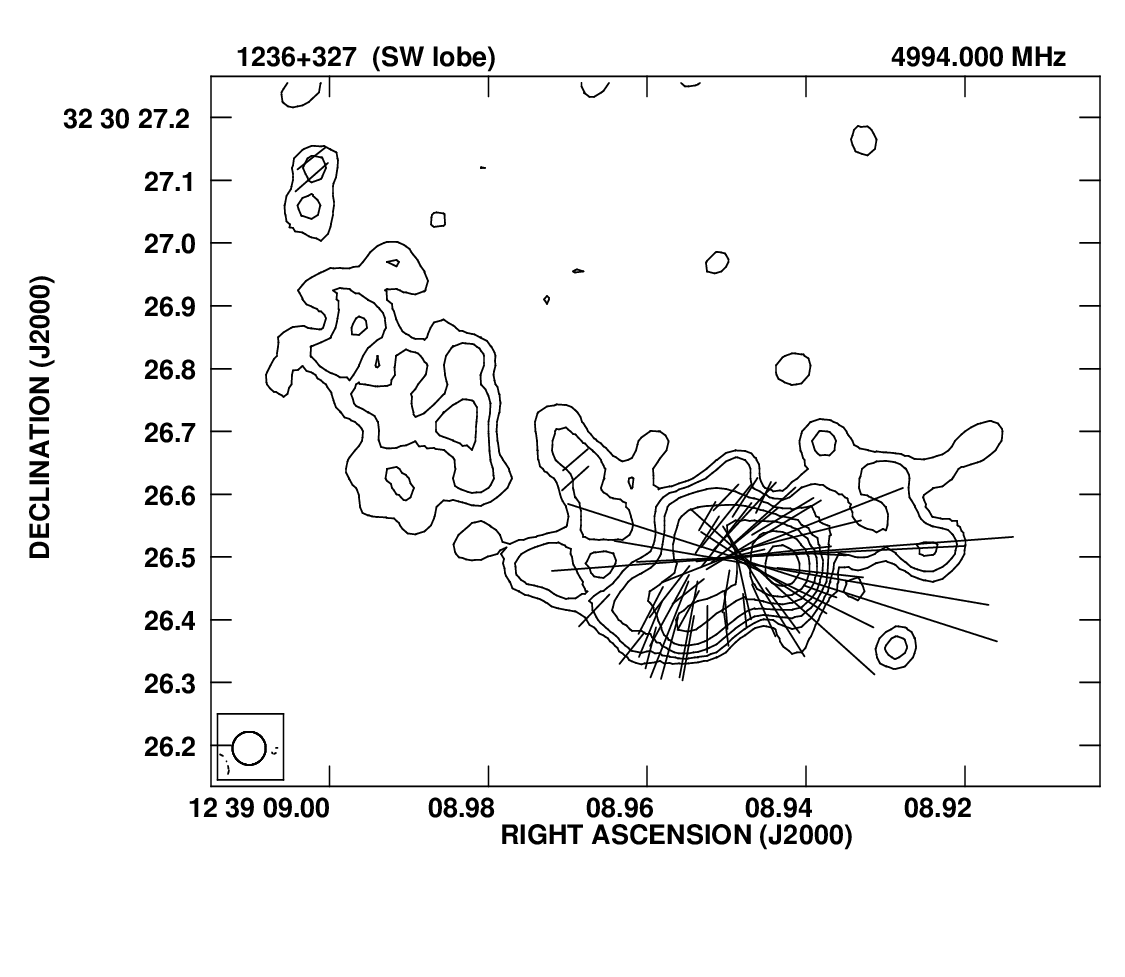}
\caption{MERLIN map of 1236+327 resulting from 12-h observation at 5\,GHz
(upper left). Combined MERLIN and VLA polarisation map of 1236+327 at 5\,GHz
(upper right). Lower panels show enlarged lobes of MERLIN-only map.
Contours increase
by a factor 2 and the first contour level corresponds to $\approx 3\sigma$.
Polarisation line of $1\arcsec$ amounts to 3.33\,mJy/beam for all maps.}
\label{1236+327_MERLIN_5GHz.map}
\end{figure*}

\begin{table*}[t]
\caption[]{Flux densities of sources principal components at 1.7\,GHz and 4.9\,GHz}
\begin{center}
\begin{tabular}{@{}c c c r r c r c c c r@{}}
\hline
\hline
   & & & & & & & & \\
~~~Source & R.A. & Dec. & \multicolumn{1}{c}{${\rm S_{1.7 GHz}}$} & \multicolumn{1}{c}{${\rm S_{4.9 GHz}}$} & $\alpha_{1.7\mathrm{GHz}}^{4.9\mathrm{GHz}}$ & ${\rm S_{8.4 GHz}}$ & $\alpha_{4.9\mathrm{GHz}}^{8.4\mathrm{GHz}}$ & $\theta_{1}$ & $\theta_{2}$ & PA~~~~~~\\
~~~Name   & h~m~s & $\degr$~$\arcmin$~$\arcsec$ & \multicolumn{1}{c}{mJy} & \multicolumn{1}{c}{mJy}& &
\multicolumn{1}{c}{mJy}& &$\arcsec$ &$\arcsec$ & $\degr$~~~~~~\\
~~~(1)    & (2) & (3)   & \multicolumn{1}{c}{(4)} & \multicolumn{1}{c}{(5)} & (6) & \multicolumn{1}{c}{(7)} &(8) & (9)& (10)& (11)~~~~~~\\
\hline
\hline
~~~1009+408& 10 12 51.83& 40 39 05.74& 154.81&
           85.39&$-$0.55&51.87&$-$0.90&0.28&0.23&162~~~~~~\\
        & 10 12 51.82& 40 39 04.18& ---  & 6.44 &---&---&---&---&---&---~~~~~~\\
        & 10 12 51.74& 40 39 01.82& 186.33& 93.45&$-$0.64&57.87&$-$0.88&0.52&0.36&20~~~~~~\\
\hline
~~~1125+327& 11 28 02.61& 32 30 50.16& 83.68 &
68.59&$-$0.18&---&---&0.46&0.30&8~~~~~~\\
        & 11 28 02.44& 32 30 46.54& ---   &  0.99& --- &--- &
---&---&---&---~~~~~~\\
        & 11 28 02.35& 32 30 44.95& 115.58& 76.77&$-$0.38&---&---&0.58&0.41&34~~~~~~\\
        & 11 28 02.32& 32 30 44.47& 53.80 & 32.30&$-$0.47&---&---&0.38&0.32&64~~~~~~\\
\hline
~~~1236+327& 12 39 09.18& 32 30 27.67& 413.42&
             138.84&$-$1.01&---&---&0.47&0.29&93~~~~~~\\
        & 12 39 09.06& 32 30 27.42& 5.01  & --- &---&---&---&0.18&0.10&19~~~~~~\\
        & 12 39 09.00& 32 30 26.94& 122.41& --- &---&---&---&0.74&0.25&47~~~~~~\\
        & 12 39 08.95& 32 30 26.51& 166.90& 80.64&$-$0.68&---&---&0.31&0.16&82~~~~~~\\
\hline
~~~1334+417& 13 36 26.54& 41 31
11.54&151.45&69.09&$-$0.73&---&---&0.48&0.25&133~~~~~~\\
        & 13 36 26.52& 41 31 11.95&125.05 & --- &---&---&---&0.93&0.49&135~~~~~~\\
        & 13 36 26.20& 41 31 15.46& 58.22 &
44.12&$-$0.26&---&---&0.29&0.13&148~~~~~~\\
\hline
~~~1542+323& 15 44 48.43& 32 08 44.06&
             303.42&140.71&$-$0.71&---&---&0.37&0.19&161~~~~~~\\
        & 15 44 48.33& 32 08 46.13&
404.75&164.80&$-$0.83&---&---&0.40&0.17&152~~~~~~\\
\hline
~~~1723+406& 17 25 16.34& 40 36 42.15&97.98&42.75
           &$-$0.77&---&---&0.30&0.06&174~~~~~~\\
        & 17 25 16.33& 40 36 43.48& 322.92&79.46&$-$1.30&---&---&0.82&0.27&180~~~~~~\\
        & 17 25 16.21& 40 36 39.04& 351.39&110.64&$-$1.07&---&---&0.52&0.25&65~~~~~~\\
\hline
\end{tabular}
\end{center}
\label{table2}
{\small
Description of the columns:
\begin{itemize}
\item[---]{Col.~~(1): Source name in the IAU format;}
\item[---]{Col.~~(2): Component right ascension (J2000) as measured at
1.7\,GHz;}
\item[---]{Col.~~(3): Component declination (J2000) as measured at
1.7\,GHz;} 
\item[---]{Col.~~(4): MERLIN flux density (mJy) at 1.7\,GHz obtained using
JMFIT;} 
\item[---]{Col.~~(5): VLA flux density (mJy) at 4.9\,GHz obtained using
JMFIT;} 
\item[---]{Col.~~(6): Spectral index between 1.7 and 4.9\,GHz calculated using
flux densities in columns (4) and (5);}
\item[---]{Col.~~(7): VLA flux density (mJy) at 8.4\,GHz obtained using
JMFIT;} 
\item[---]{Col.~~(8): Spectral index between 4.9 and 8.4\,GHz calculated
using flux densities in columns (5) and (7);}
\item[---]{Col.~~(9): Deconvolved component major axis angular size at 1.7\,GHz
obtained using JMFIT;}
\item[---]{Col.~(10): Deconvolved component minor axis angular size at 1.7\,GHz
obtained using JMFIT;}
\item[---]{Col.~(11): Deconvolved major axis position angle at 1.7\,GHz
obtained using JMFIT;}
\end{itemize}
}
\end{table*}

\begin{figure*}
\centering
\includegraphics[height=8cm,width=8cm]{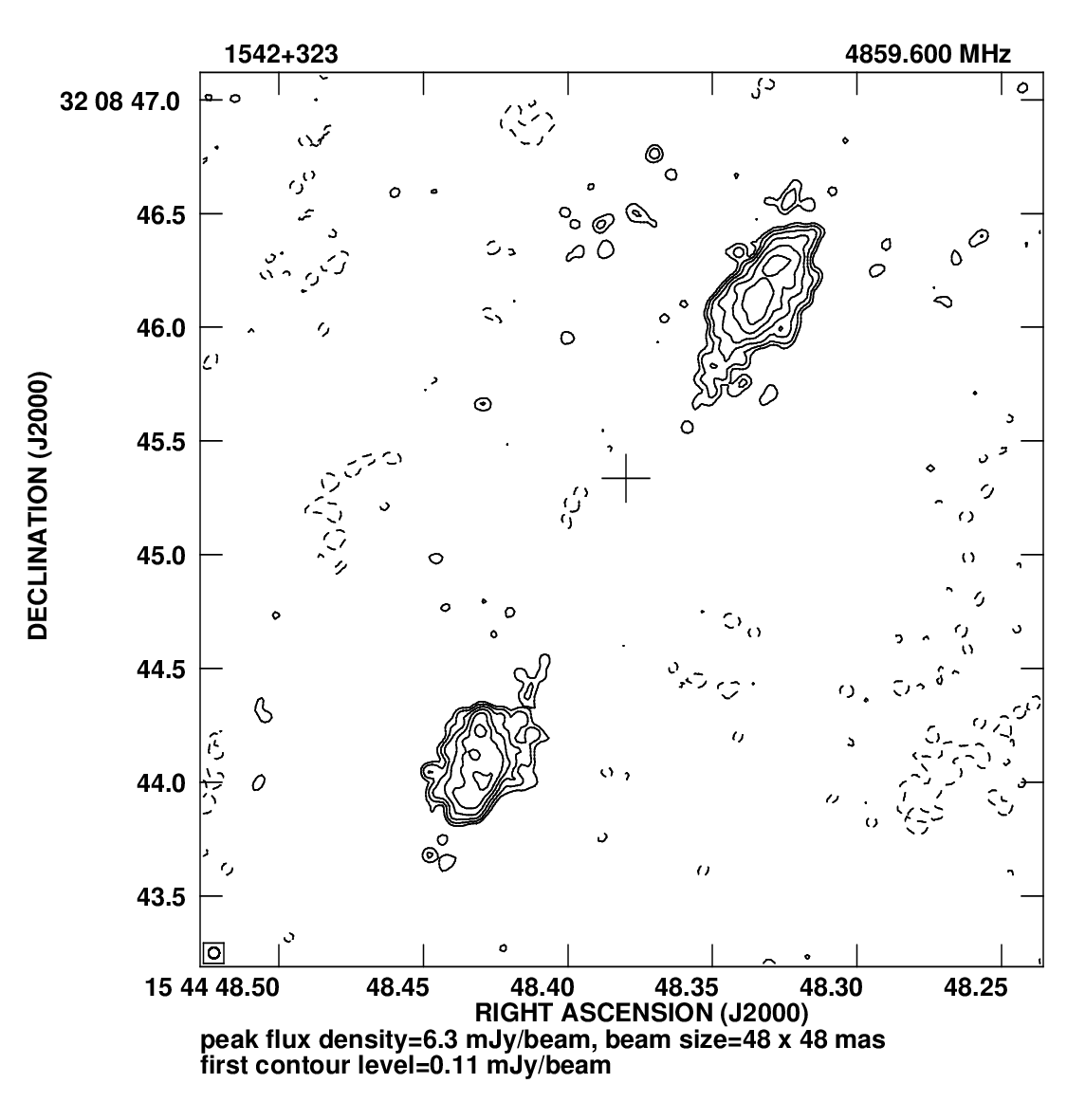}
\includegraphics[height=8cm,width=8cm]{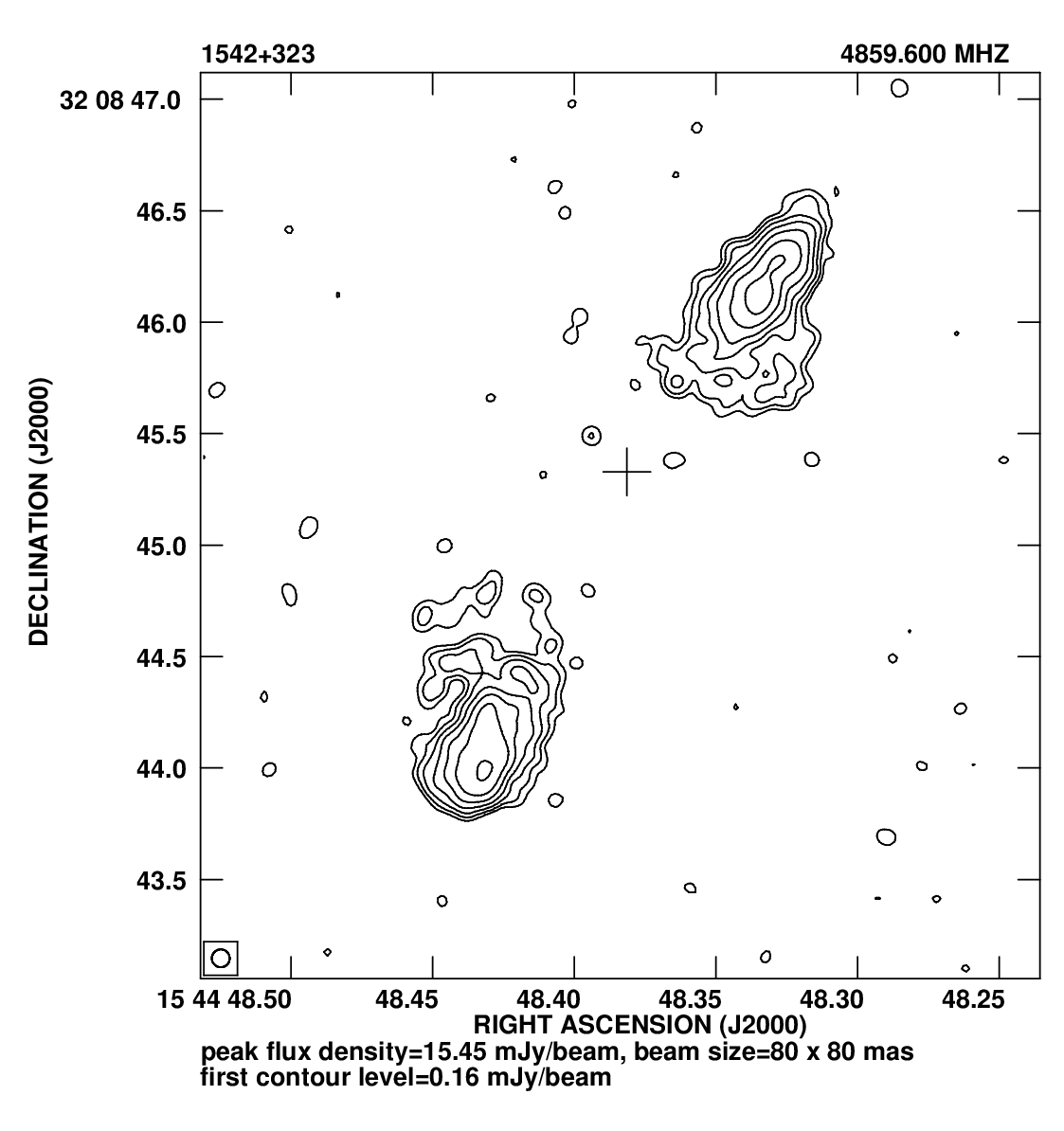}
\includegraphics[height=8cm,width=8cm]{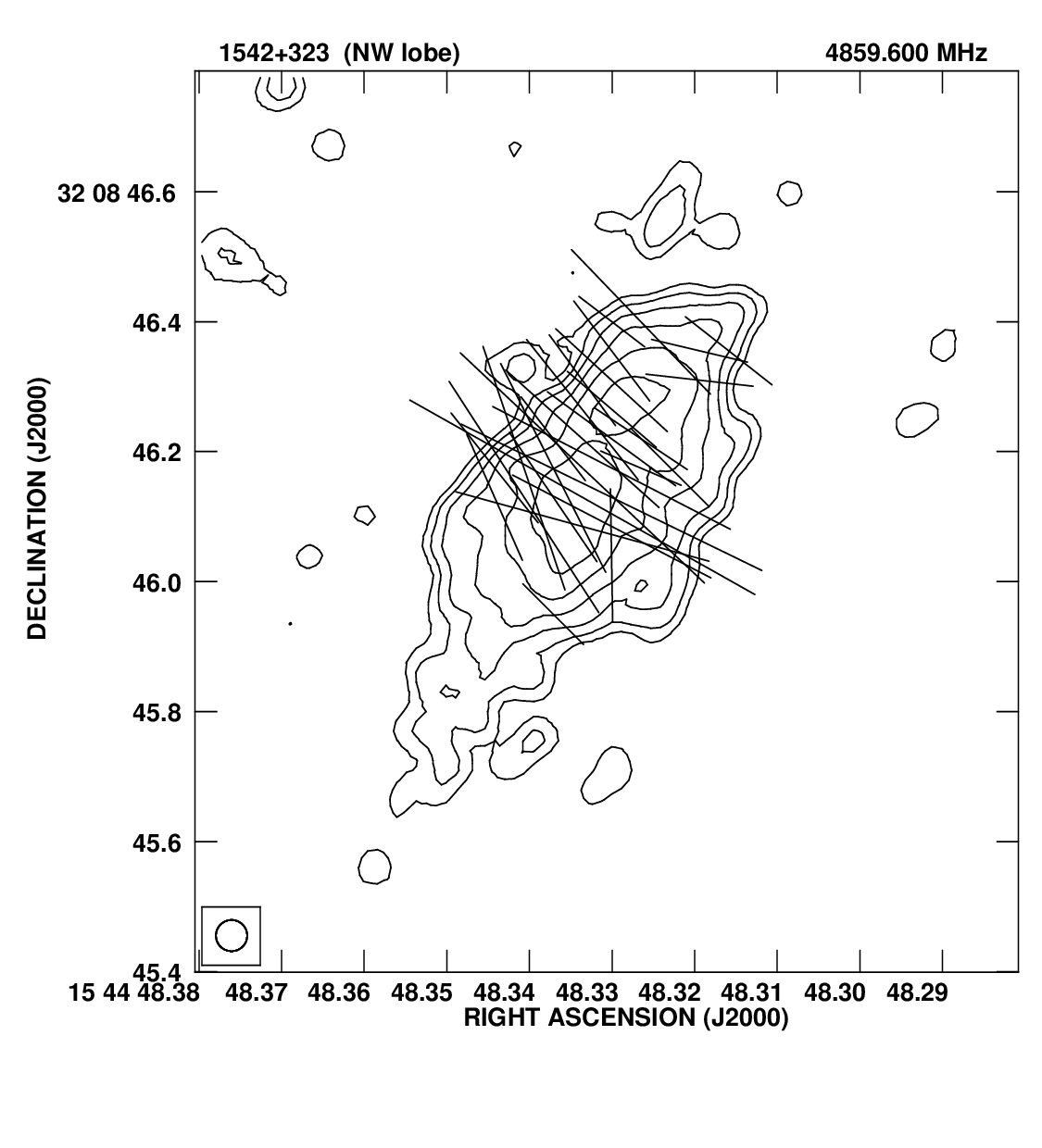}
\includegraphics[height=8cm,width=8cm]{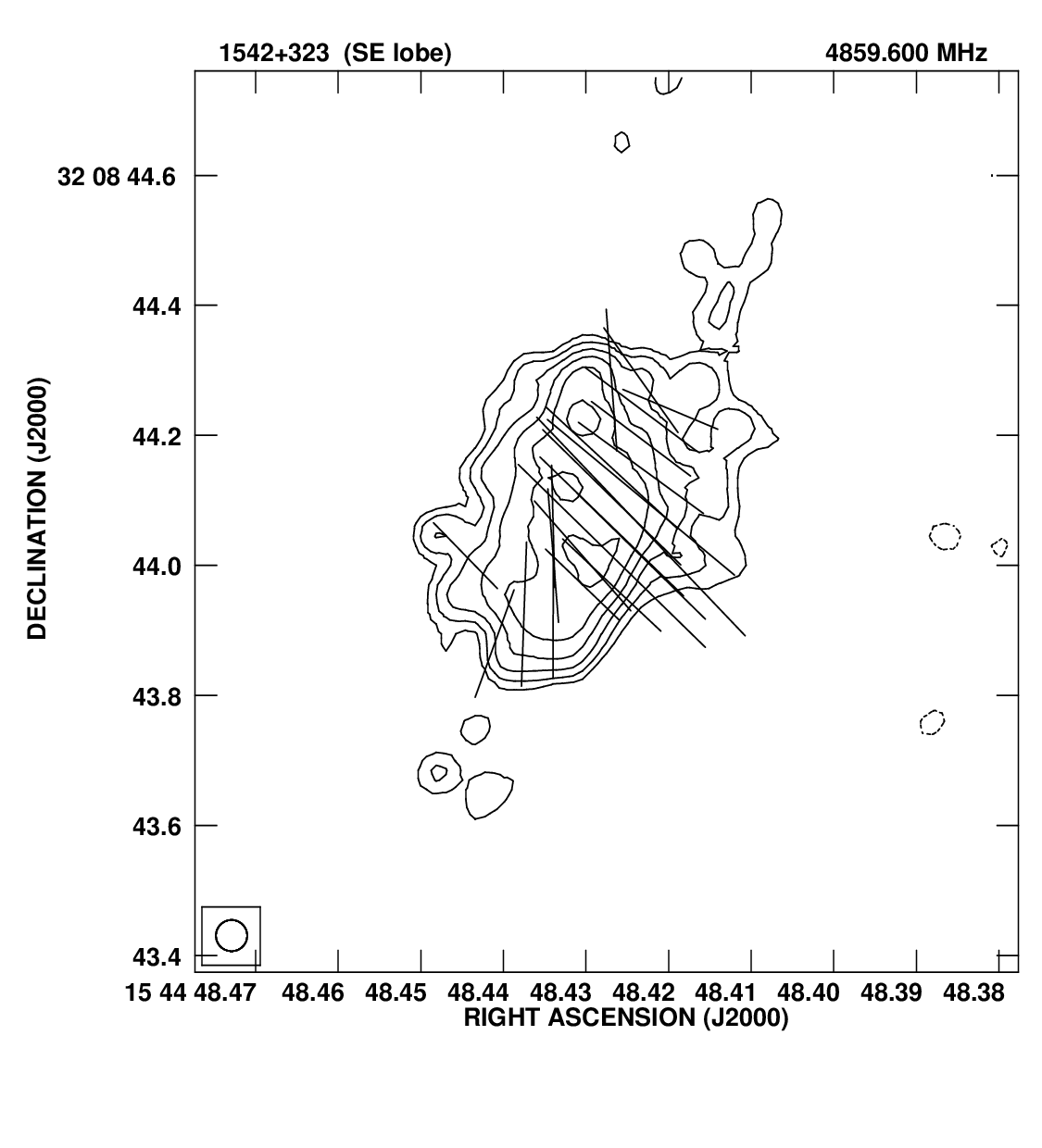}
\caption{MERLIN polarisation map of 1542+323 resulting from 12-h
observation at 4.9\,GHz (upper left). 
Combined MERLIN and VLA polarisation map of 1542+323 at 4.9\,GHz (upper right).
Lower panels show enlarged lobes of MERLIN-only map.
Contours increase by a factor 2 and
the first contour level corresponds to $\approx 3\sigma$.
Polarisation line of $1\arcsec$ amounts to 1.33\,mJy/beam for all maps.
Crosses indicate the position of an optical object found using the SDSS/DR3.}
\label{1542+323_MERLIN_5GHz.map}
\end{figure*}

The positions of the optical identifications of the sources, if available,
extracted either from the POSS plates using the Automatic
Plate Measuring (APM) machine or Third Data Release of the
Sloan Digital Sky Survey (SDSS/DR3) are marked with crosses on the radio images.

In addition to the observations described above, an unpublished 8.4-GHz VLA 
observation of 1009+408 made by \citet{jvas1} as part of the Jodrell Bank--VLA
Astrometric Survey (JVAS) has been included here as this may be helpful in
the presentation of our ideas (Fig.~2a 
--- lower panel; Table~\ref{table2}).

The basic properties of the 6~sources are given in Table~\ref{table1}.
Except for 1125+327 whose redshift estimate and linear size are
commented upon in Section 4, the redshifts of the sources are unknown at the
present time.
The estimated total fluxes at 1.7 GHz quoted in Table 1 have been derived from
an interpolation between the 1.4 GHz and 4.85 GHz fluxes assuming
a constant spectral index.   These values can provide an assessment
of the missing flux in the MERLIN 1.7 GHz images when compared
with a summation of the fluxes of the main components in the MERLIN images.

The flux densities of the components (or well defined features) of the
sources at both 1.7 GHz and 4.9 GHz were measured using the AIPS task JMFIT 
and their spectral indices
calculated. The results are listed in Table~\ref{table2}.

\section{Comments on individual sources}\label{s-comments}

{\bf\object{1009+408}}. MERLIN and VLA images of this double source are shown
in Figs.~\ref{MERLIN_maps} and~2a. 
The more extended emission between the two main
lobes seen in the 4.9-GHz VLA image is barely visible in the MERLIN 1.7-GHz
image probably because of a lack of short spacings in the MERLIN data at 
1.7\,GHz. The only indication of a core in any of the images is in the 4.9-GHz
VLA image, in which there is a peak of emission
at RA=10$^h$12$^m$51.$^s$82, Dec.=+40\degr39\arcmin04\farcs18, although even
this is very doubtful as there is no indication of a core at this position
in the MERLIN 5-GHz image. However,
its determined Gaussian model parameters are given in
Table~\ref{table2}. Crosses on the radio images indicate the position
of a galaxy found using the SDSS/DR3. 
The northern lobe does not show any very compact features in any of the
images and its spectral index is moderately steep.
There is a hotspot at the edge of the southern lobe in both the 5-GHz MERLIN 
images and in the 8.4-GHz VLA image.
There is also an asymmetry in polarisation between the lobes
(Fig.~2a). 

{\bf\object{1125+327}}. MERLIN and VLA images of this double source
are shown in Figs.~\ref{MERLIN_maps} and~2b. 
A published VLA image of this radio galaxy at 1.46\,GHz 
by \citet{mc83} showed it to have a triple structure with two bright components 
and one very weak one to the south-west of those two. As our MERLIN images at a 
similar frequency showed only a double structure with no south-western
component, the raw 1.46-GHz data for this 
source originally used by \citet{mc83} was extracted from the VLA archive 
and reprocessed. No third component to the south-west was found and so it is 
assumed that this component in the original 1.46-GHz VLA image is an artefact. 

There is an intensity peak in the 4.9-GHz VLA image at 
RA=11$^h$28$^m$02.$^s$44, Dec.=+32\degr30\arcmin46\farcs54
(Table~\ref{table2}), which may be a radio core. There is no such feature in 
the 1.7-GHz MERLIN image (Fig.~2b) 
so the core could have a 
flat or inverted spectrum as one might expect. Crosses on the radio
images indicates the position of a galaxy found on the POSS plates using the 
APM. \citet{ma98} published a photometric redshift for this object:
$z=0.75$. However, it is to be noted that the position listed by \citet{ma98} 
is somewhat different from that in POSS which, together with the limited
accuracy of the method, makes the redshift determination rather insecure.
Bearing this in mind, but assuming the above quoted redshift, a linear
extent of $27.6 h^{-1}$\,kpc has been calculated which slightly exceeds the
formal limit of the MSO class.

The extended emission visible in the VLA 4.9-GHz image is resolved out by 
MERLIN and consequently a large fraction (50\%) of the flux in the
1.7-GHz image is missing.
Also, $\sim$10\% of the flux in the 4.9-GHz VLA image has not been
accounted for in the process of fitting the gaussian components listed
in Table~\ref{table2} --- this missing flux can be attributed to the diffuse 
bridge-like part of the NE lobe.

The southern lobe has two small peaks, one of which seems to be a hotspot as 
indicated by the 5-GHz MERLIN image (Fig.~\ref{MERLIN_maps}). The NE
lobe is also edge-brightened by a flat spectrum hotspot.

{\bf\object{1236+327}}. Following the analysis of the initial MERLIN and
VLA observations of 1236+327 (Figs.~\ref{MERLIN_maps}
and~2c), 
it was thought that the
source might have a mixed morphology with the SW end clearly in an
`active' phase, but the NE end possibly fading out. Similar cases
of such sources have been described by \citet{lon83}. They investigated 
sources
with asymmetric structures and their research revealed that
in each source, one hotspot is more compact and of less steep spectrum
than its counterpart on the opposite side of the source. They divided
hotspots into two different classes and postulated that class II hotspots
are aged and expanded versions of class I hotspots. If the energy supply to
a class I hotspot is removed, a steepening of the spectrum, an increase in
size and a possible tangling of the magnetic field are to be expected. 
If there is a `working' jet supplying the lobe, the magnetic field lines 
have a tendency to lie perpendicular to the source axis.

In order to determine more fully the structure of 1236+327, a full-track
MERLIN observation at 5\,GHz was made. The resulting image is shown
in Fig.~\ref{1236+327_MERLIN_5GHz.map} with the lower panels showing
the polarised emission of the two lobes of the source in more detail.
This full-track 5-GHz MERLIN data were also combined 
with the existing VLA 4.9-GHz data to produce the image also shown in
Fig.~\ref{1236+327_MERLIN_5GHz.map} (upper right panel).
It can be seen that the SW lobe has a comparatively simple, classical
FR\,II type of structure, whereas the
NE lobe appears to be very complex with several possible hotspots.
It is to be noted that one can draw an almost perfect straight line between
the western hotspot in the NE lobe, the peak of emission located
at RA=12$^h$39$^m$09\fs13, Dec.=+32\degr30\arcmin27\farcs6 and the
hotspot in the SW lobe, indicating that this peak of emission
between the two lobes is perhaps the core. Assuming this to be the case
means that the source has a highly asymmetric arm ratio for its two lobes
($\sim$5.4), which is even larger than the value of $\sim$5 quoted for
3C459 by \citet{tsm03}. 

\citet{best95} showed that the structures
of FR\,II sources are probably determined by relativistic, environmental and
intrinsic asymmetries. According to \citet{ars00} relativistic effects
are more important for quasars and high luminosities, whereas
intrinsic/environmental asymmetries are more important for low luminosities
and radio galaxies, and are more significant on small physical scales.
The asymmetry in the structure of 1236+327 is probably caused by both 
orientation on the sky and interaction with the surrounding medium whereas
the conjecture that this source might have a mixed morphology with one lobe in
an active phase, and another one possibly fading \citep{lon83} is ruled out. 

{\bf\object{1334+417}}. This source has a classical double structure 
with two
edge-brightened radio lobes (Figs.~\ref{MERLIN_maps} and~2d).
The southern lobe has a steep spectrum and the polarisation is detectable
only in this lobe. 
The bridge of emission between the two lobes, clearly visible in the
VLA 4.9\,GHz image, is resolved out in the MERLIN 1.7\,GHz image.
As the flux emitted by the bridge contributes significantly to
the total flux at both frequencies, the sum of the fluxes of
the fitted gaussians listed in Table~\ref{table2} is far less than the total
flux indicated in Table~\ref{table1}. Overall, the source has a typical
FR\,II-like morphology i.e. its lobes are clearly edge-brightened.

{\bf\object{1542+323}}. The initial MERLIN and VLA images
(Figs.~\ref{MERLIN_maps} and~2e) 
indicated that this source might indeed be a fader. It did not appear to
have a core, nor did the lobes, which were not edge-brightened, appear to
have hotspots (Fig.~2e). 

A full-track MERLIN observation at 4.86\,GHz, the same frequency as the 
VLA observation, was made in order
to more fully investigate the source lobe structure and to search for a
core. The resulting images
of the overall structure of the source and the individual lobes including
polarisation vectors on an enlarged scale are shown in
Fig.~\ref{1542+323_MERLIN_5GHz.map}. There is still no indication of a
core or hotspots. It is to be noted that there is no change in the direction
of the polarisation vectors (i.e. magnetic field) in the regions of maximum
intensity, which one might expect if they were active hotspots. It would
appear that this source could be a fader.

{\bf\object{1723+406}}. MERLIN and VLA images of this double source
are shown in Figs.~\ref{MERLIN_maps} and~2f. 
This is clearly a triple source with an interesting bent structure in all
our images. There appears to be a core, part of a jet and a radio lobe,
although all these three components have very steep spectra, which possibly
indicates that the true core is still hidden. Our 5-GHz MERLIN observations 
(Fig.~\ref{MERLIN_maps}) resolved out the extended emission visible in the
4.9-GHz VLA image and in the 1.7-GHz MERLIN image (Fig.~2f)
so only two compact components show up: a core and probably a hotspot from the
southern lobe and some diffuse emission from the jet. The radio
structure of the source on the two sides of the nucleus exhibits different
Fanaroff-Riley morphologies. Sources of that kind are called HYbrid
MOrphology Radio Sources \citep[HYMORS;][]{gw2000}.

\section{Discussion}

\citet{feretti84} found a significant correlation between the core
luminosity of radio galaxies $P_c$ and the total radio power $P_t$
(core\,+\,extended structure), but there was quite a large dispersion in this
correlation. Also, over 30\% of their sample only had upper limits for the
core luminosities, which could have been due to a lack of observational 
sensitivity. Consequently,
\citet{giov88} made much greater sensitivity observations with the VLA at
4.9\,GHz of those sources whose cores had not been detected. A conclusion of
their work was that two of their sources, whose cores were still undetected
at a level which was at least two sigma below the expected $P_c$~---~$P_t$
correlation, could represent `dying' sources, i.e. the sources in which the
supply of energy from the nucleus had ceased. A further four sources with
$P_c$~---~$P_t$ approximately one sigma below the expected correlation were
also probably `dying' sources.

The MERLIN 5-GHz image of 1542+323 (Fig.~\ref{1542+323_MERLIN_5GHz.map}) shows
two radio lobes without visible hotspots and without a core,
the upper limit to the flux density of the latter being below that 
expected from the correlation of core luminosity at 4.9\,GHz ($\log P_c$) vs.
total luminosity at 408\,MHz ($\log P_t$) for all but the most powerful of
radio galaxies (\citet{giov88} --- equation~1). (The source
total flux density at 408\,MHz (1.99\,Jy) was estimated using a
spectral index between 1.4\,GHz and 365\,MHz of $-0.69$, the latter being
calculated from the 1.4-GHz flux given in Table~\ref{table1} and the flux
density at 365\,MHz quoted in the Texas catalogue \citep{do96}.)
Thus, it would appear that 1542+323 is possibly a CSS/MSO fader.
However, although the above indicates that there is no longer a power supply
from a central engine, the polarised emission in the northern lobe, in which
there would appear to be no or very little Faraday rotation and depolarisation,
indicates a measure of doubt about this.

The flux density ratios
of the lobes of sources in the sample have been determined to see if
there are any asymmetries, as it has already been noticed that in MSOs
large lobe flux density asymmetries sometimes exist. It has also been
postulated that most CSS sources may pass through a phase of evolution in 
an asymmetric
environment \citep{sa01, sa02}. There is only a small asymmetry in the
flux densities of the lobes of our sources at the level of $\sim$1.3.

A number of theoretical predictions of the existence of a class of young
sources which never grow to large sizes have appeared recently. For example,
\citet{ghis04} have suggested a model of aborted jets. In this model, the jets
in a radio-quiet object could be accelerated to relativistic speeds and the
object would thus become radio-loud. Eventually, when the active period comes
to an end, diffuse, weak radio lobes without visible jets should be seen.

In interpreting the results of our observations, and 1542+323 in
particular, the scenario proposed by RB97 could be adopted, namely
that the energy supply from the jets might be interrupted. 
Taking into consideration the predictions made by RB97, it would appear
that the evolutionary track of RLAGNs proposed by
\citet{sn99, sn00} is not necessarily the only one. In fact, they
themselves admit that it is unclear whether all young sources evolve into
LSOs, and their plot shows an additional branch indicating sources 
which have left the `main sequence' at an earlier stage. 
(They labelled them `drop-outs'). Thus, a whole family (a continuum?)
of such tracks might exist and the one shown by
\citet{sn99} appears as the {\em only} one simply because of a selection
effect related to the Malmquist bias.

Therefore, the following scenario might be considered. Suppose the energy
supply from the central engine cuts off after e.g. $\sim$$10^5$ years i.e. a
timescale typical for the spectral ages of CSS/MSOs \citep{murgia99}. Under
such circumstances a `dying' CSS source would result which, provided it is
not beamed towards us, would be perceived as an MSO. Just as for large-scale
faders, a `dying' MSO, unlike a standard, high radio luminosity MSO, should
be relatively weak because of a lack of fuelling and should not be
edge-brightened because there are no (or hardly any) jets pushing through
the intergalactic medium. Thus, the hotspots should have faded away. What
should be seen therefore is nothing more than diffuse lobes without well
defined hotspots analogous to those observed in large-scale faders.

The major problem with accepting the above scenario is that LSO faders
should have very steep spectra \citep[cf.][]{kom94}, which is not the case
for 1542+323: it has a mean spectral index of $\alpha=-0.78$ --- see
Tables~\ref{table1} and~\ref{table2}. However, if it is assumed that
expansion losses dominate in subgalactic-scale faders and that they could
occur in a comparatively short period of time --- $10^4$\,years (J.P. Leahy,
priv. comm., but see also RB97), then the morphological signs of fading and
a decrease in luminosity would take precedence over the transformation of
the spectrum into a very steep one. Thus, the lobes would quickly take the
typical form of a fader without their spectra showing signs of ageing for
frequencies below 5\,GHz. Thus, 1542+323 could still be labelled as a
`fader'.

Finally, the issue of the `unification' of small and large-scale faders
should be discussed. It appears fairly obvious that a single mechanism
should be at work in faders regardless of their sizes. The simplest reason
for cessation of activity of an AGN is that there is no more matter to be
accreted onto a Supermassive Black Hole (SMBH) in a host galaxy centre. An
interesting alternative to that scheme, namely the mechanism of
thermal-viscous instabilities in the accretion disk of an AGN, was proposed
by \citet{hse01} --- hereafter HSE \citep[see also][]{siem96, siem97,
jcss04}. According to their model, galaxies spend the greater part of their
lifetime ($\sim$70\%) in a `quiescent' state and $\sim$30\% in an active
state. RB97 do not investigate the physical mechanism that drives the short
and frequent outbursts of activity in RLAGNs, but as suggested by
\citet{B99} their model can be combined with that of HSE who predict that,
since the length of the active phase of an AGN as well as the timescale of
activity re-occurrence is determined by the square of the mass of the SMBH,
it follows that the length of an active phase {\em can} be as short as
$10^5$ years which, as indicated above, is typical for MSOs. Based upon the
relationship given by HSE, this corresponds to a SMBH mass of $\sim$$10^8
M_{\sun}$.  It is to be noted that, according to \citet{osh02} and
\citet{wu02}, the range of SMBH masses residing in the centres of RLAGNs is
very wide: $2\cdot 10^6 M_{\sun}$ -- $7\cdot 10^9\,M_{\sun}$. This
corresponds to a range of 7~orders of magnitude for allowable timescales for
the duration of an active phase. Thus, the theory of thermal-viscous
instabilities in the accretion disks of AGNs can be used to explain the
existence of a very wide range of timescales of activity periods in RLAGNs.
Eventually, this translates to a wide range of linear sizes starting from
CSOs through MSOs up to the very extended, old LSOs which, if their sizes
exceed 1\,Mpc, are termed Giant Radio Galaxies \citep[see e.g.][]{lara01}.

\section{Summary and conclusions}

The compactness of MSOs is interpreted as a direct consequence of
their `youth'. They are believed to be the precursors of
larger/older objects and so eventually become LSOs.
It is unclear whether {\em all} young sources actually evolve to become very
extended objects. Some of them may be short-lived phenomena and can leave
the main evolutionary track earlier because of a lack of stable fuelling.
Theoretical predictions made by RB97 combined with the model given by
HSE allow for short periods of activity provided that the mass of the central
SMBH is sufficiently low.
The main goal of this study was to search for relic sources among those with
subgalactic sizes. Finding examples of RLAGNs in which the energy transport
has switched off early so that they have become compact faders would provide
evidence that the scenario outlined above is plausible i.e. the
activity of a RLAGN could decline at {\em any} stage of its evolution.

A new sample of weak CSS sources has been established and the sources
observed to see if the evolutionary track for strong radio sources 
could be applied to weak ones. 
In this paper 6~sources from the sample which, based on a MERLIN 5-GHz
snapshot survey seemed to be good candidates for `dying' MSOs, have been
studied. Only one of them, 1542+323, possibly appears to be a fader.
The other five sources show no obvious signs of being switched off.

\begin{acknowledgements}

\item MERLIN is operated by the University of Manchester as a National
Facility on behalf of the Particle Physics \& Astronomy Research Council
(PPARC).

\item VLA is operated by the U.S. National Radio Astronomy Observatory
which is operated by Associated Universities, Inc., under cooperative
agreement with the National Science Foundation.

\item The Automatic Plate Measuring (APM) machine is a National Astro\-nomy
Facility run by the Institute of Astronomy in Cambridge (UK). Official website:
http://www.ast.cam.ac.uk/$\sim$apmcat.

\item This research has made use of the NASA/IPAC Extragalactic Database
(NED) which is operated by the Jet Propulsion Laboratory, California
Institute of Technology, under contract with the National Aeronautics and
Space Administration.

\item Use has been made of the third release of the Sloan Digital Sky
Survey (SDSS) Archive. Funding for the creation and distribution of the
SDSS Archive has been provided by the Alfred P. Sloan Foundation, the
Participating Institutions, the National Aeronautics and Space
Administration, the National Science Foundation, the U.S. Department of
Energy, the Japanese Monbukagakusho, and the Max Planck Society. The SDSS
Web site is http://www.sdss.org/. The SDSS is managed by the Astrophysical
Research Consortium (ARC) for the Participating Institutions. The
Participating Insti\-tutions are The University of Chicago, Fermilab, the
Insti\-tute for Advanced Study, the Japan Participation Group, The Johns
Hopkins University, Los Alamos National Labora\-tory, the
Max-Planck-Institute for Astronomy (MPIA), the Max-Planck-Institute for
Astrophysics (MPA), New Mexico State University, University of Pittsburgh,
Princeton University, the United States Naval Observatory, and the
University of Washington.

\item Part of this research was made when M.K.B. stayed at Jodrell Bank
Observatory and received a scholarship provided by the EU under the
Marie Curie Training Site scheme.

\item A.M. acknowledges the receipt of a travel grant funded by Radio\-Net
as a part of the Trans-National Access (TNA) programmes.

\item We thank Tom Muxlow for help with the data reduction.

\item We thank the anonymous referee for a number of valuable suggestions
leading to the improvement of the original version of the paper.

\end{acknowledgements}

\end{document}